\documentclass[fleqn,usenatbib]{mnras}

\usepackage{newtxtext,newtxmath}

\usepackage[T1]{fontenc}

\DeclareRobustCommand{\VAN}[3]{#2}
\let\VANthebibliography\thebibliography
\def\thebibliography{\DeclareRobustCommand{\VAN}[3]{##3}\VANthebibliography}

\usepackage{graphicx}	
\usepackage{amsmath}	

\usepackage{color}
\usepackage{lastpage}

\title[ELUCID-DESI I: MPI code]{ELUCID-DESI I: A Parallel MPI Implementation of the Initial Condition Solver for Large-Scale Reconstruction Simulations}

\author[Wensheng Hong et al.]{
Wensheng Hong$^{1,2}$, Xiaohu Yang$^{1,2}$\thanks{xyang@sjtu.edu.cn}, Junde Li$^{1,2}$, Huiyuan Wang$^{3}$, Zhao Chen$^{1,2}$, Hong-Ming Zhu$^{4}$, \newauthor ~Qingyang Li$^{1,2}$, Yizhou Gu$^{1,2}$, Youcai Zhang$^{5}$, Feng Shi$^{6}$, Jiaxin Han$^{1,2}$, Yu Yu$^{1,2}$, Zhongxu Zhai$^{1,2}$
\\
$^{1}$State Key Laboratory of Dark Matter Physics, Tsung-Dao Lee Institute \& School of Physics and Astronomy, Shanghai Jiao Tong University, Shanghai 201210, China
\\
$^{2}$Shanghai Key Laboratory for Particle Physics and Cosmology, and Key Laboratory for Particle Physics, Astrophysics and Cosmology, Ministry of Education, \\~~Shanghai Jiao Tong University, Shanghai 200240, China
\\
$^{3}$Key Laboratory for Research in Galaxies and Cosmology, Department of Astronomy, University of Science and Technology of China, Hefei, Anhui 230026, China
\\
$^{4}$National Astronomical Observatories, Chinese Academy of Sciences, 20A Datun Road, Beijing 100101, China\\
$^{5}$Shanghai Astronomical Observatory, Nandan Road 80, Shanghai 200030, China \\ 
$^{6}$School of Aerospace Science And Technology, Xidian University, Xi'an 710126, China
}
\date{Accepted XXX. Received YYY; in original form ZZZ}

\pubyear{\the\year{}}

\begin{document}
\label{firstpage}
\pagerange{\pageref{firstpage}--\pageref{lastpage}}
\maketitle

\begin{abstract}
We present a highly scalable, MPI-parallelized framework for reconstructing the initial cosmic density field, designed to meet the computational demands of next-generation cosmological simulations, particularly the upcoming ELUCID-DESI simulation based on DESI BGS data. Building upon the Hamiltonian Monte Carlo approach and the FastPM solver, our code employs domain decomposition to efficiently distribute memory between nodes. 
Although communication overhead increases the per-step runtime of the MPI version by roughly a factor of eight relative to the shared-memory implementation, our scaling tests-spanning different particle numbers, core counts, and node layouts-show nearly linear scaling with respect to both the number of particles and the number of CPU cores. Furthermore, to significantly reduce computational costs during the initial burn-in phase, we introduce a novel ``guess'' module that rapidly generates a high-quality initial density field. 
The results of the simulation test confirm substantial efficiency gains: for $256^3$ particles, 53 steps ($\sim$ 54 core hours) are saved, accelerating convergence by a factor of $\sim$ 18; for $1024^3$, 106 steps ($\sim$7500 core hours), achieving a speedup factor of $\sim$ 3. The total core hour gain grows with the number of particles, rendering large-volume reconstructions computationally practical for upcoming surveys, including our planned ELUCID-DESI reconstruction simulation with $4096^3$ particles. We estimate that achieving convergence for this scale (targeting DESI-BGS data) requires about 800 HMCMC steps ($\sim$ 5 million core hours). Our initial guess module will save approximately 360 steps ($\sim$2.3 million core hours), reducing the total computational time by about 45\%.
\end{abstract}




\section{Introduction}

We are entering a new era of observations powered by fourth-generation wide-field spectroscopic surveys. The Dark Energy Spectroscopic Instrument \citep{DESI2016a,DESI2016b}, the Prime Focus Spectrograph \citep{Takada_2014}, the Chinese Space Station Survey Telescope \citep{CSST2011,CSST2019,CSST2025}, and the Jiao-tong University Spectroscopic Telescope \citep{JUST_2024} will collectively map hundreds of millions of galaxies over cosmological volumes that exceed those of earlier surveys by roughly an order of magnitude. In addition to their central objectives of probing the nature of dark energy and the properties of dark matter, these surveys will provide a powerful resource for studying how galaxies with a wide range of characteristics form and evolve in concert with the build-up of cosmic dark matter structures \citep{Mo_2010}.

To understand how galaxies assemble throughout the cosmic density field, it is essential to investigate the spatial distributions and internal characteristics of both galaxies and the intergalactic medium (IGM), as well as their mutual interactions, within dark matter halos that cover a wide range of masses and reside in varied environments of the cosmic web. Traditionally, constraints on models of galaxy formation have largely come from statistical comparisons  \citep[e.g.,][]{Kauffmann_1993, Guo_2011}, assessing how well theoretical predictions match key observables such as the galaxy luminosity and stellar mass functions, galaxy clustering \citep[e.g.,][]{Zehavi_2005, Zehavi_2011, Yang_2012}, halo occupation statistics \citep[e.g.,][]{Jing_1998, Zheng_2005}, and the conditional luminosity and stellar mass functions \citep[e.g.,][]{Yang_2008, Yang_2009, WangYR_2024}. Yet, because galaxy formation involves highly complex and intertwined physical processes, these statistical approaches have not fully exploited the richness of observational data, as many processes remain effectively decoupled in such analyses. A particularly powerful way to harness observational information is through constrained simulations, in which the initial density field is directly inferred from observations. This approach enables detailed, object-by-object comparisons and thus provides a more stringent means of constraining galaxy formation physics.

In this context, considerable work has gone into developing methods to infer the initial conditions of structure formation in the nearby Universe from galaxy surveys and/or measurements of peculiar velocities \citep{Sousa_2007, Jasche_2013, Seljak_2017}. \citet{Hoffman_1991} presented a method for generating Gaussian random fields that satisfy an imposed set of constraints \citep[see also][]{Bertschinger_1987, van_de_Weygaert_1996, Kitaura_2008}. The measured peculiar motions of galaxies in the local volume have similarly been widely employed to set up initial conditions for constrained simulations \citep[e.g.,][]{Kravtsov_2002, Klypin_2003, Doumler_2013}. Alongside these backward-modeling approaches, Hamiltonian Markov Chain Monte Carlo (HMC) reconstruction techniques have been steadily improved and extended \citep{Jasche_2013, Kitaura_2013, Wang_2013, Wang_2014}.

Within this family of forward-modeling approaches, the particle-mesh (PM)–based HMC framework developed by \citet{Wang_2013, Wang_2014} (see also \citet{Jasche_2019} for a related PM+HMC work) currently delivers the most accurate reconstructions, primarily because PM itself is a mature and well-tested simulation technique. In this approach, the negative log-posterior is treated as the potential energy of a Hamiltonian system, enabling HMC to efficiently probe the extremely high-dimensional parameter space characterizing the initial density field. Each HMC step entails the evaluation of the likelihood gradient, which in practice is equivalent to executing a full cosmological simulation. This rigorous combination of statistical mechanics with cosmological dynamics has established a new standard for statistically robust, full-volume reconstruction methodologies.

Using the group catalogs \citep{Yang_2007} constructed from the Sloan Digital Sky Survey \citep[SDSS,][]{SDSS_2000} main galaxy sample DR7 and the associated value-added galaxy catalogs \citep{Blanton_2005}, we performed the ELUCID (Exploring the Local Universe with reConstructed Initial Density field) simulations \citep[e.g.][]{Wang_2014, Tweed_2017, Wang_2016}. These simulations enable us to build a one-to-one correspondence between observed galaxies and dark matter subhalos within the simulated volume \citep[e.g.][]{Yang_2018}. With this mapping in hand, we can examine how the morphology and orientation of observed galaxies are affected by the assembly histories of their host (sub)halos and by their positions in the cosmic web \citep[e.g.][]{Zhang_2021, Zhang_2022, Zhang_2025}. Moreover, it provides the observational environmental framework required to quantify how galaxy formation—covering both star formation and quenching—is governed by {\it in situ} and {\it ex situ} processes \citep[][]{Wang_2018}, among other factors.



Motivated by the opportunity to carry out much larger-volume, higher-resolution re-simulations of the local universe using the upcoming DESI BGS DR3 data \citep[e.g.,][]{Myers_2023, Hahn_2023}, we present the ELUCID-DESI project, which extends the redshift range to $z<0.4$, beyond the original ELUCID limit of $z<0.12$. Realizing this objective requires, as a crucial first step, reconstructing the initial conditions of our local universe over a volume roughly 64 times larger than before. However, the remarkable strengths of the HMC framework come at the price of very high computational demands. The key bottleneck is intrinsic: each likelihood evaluation necessitates a full forward simulation. Consequently, obtaining a well-converged Markov chain for a reconstruction over such a large volume can consume hundreds of millions of CPU core-hours, rendering routine use impractical. Furthermore, the original ELUCID configuration, like other HMC-based approaches, relies on shared-memory (OpenMP) parallelization. This architecture enforces a strict memory ceiling, since the entire problem must reside in the RAM of a single compute node—an especially restrictive limitation given the scale of upcoming datasets.

As the first paper in the ELUCID-DESI series, this work represents a step toward overcoming these challenges. We present a next-generation, massively parallel implementation of the Bayesian framework for reconstructing initial conditions. Our scheme preserves the rigorous HMC formulation of \citet{Wang_2014}, but is fundamentally re-engineered to confront both memory and computational-throughput limitations in giga- and tera-particle regimes. Our main advances are: (I) A distributed-memory, MPI-parallel implementation employing 3D domain decomposition, built upon the scalable FastPM particle-mesh solver. This removes the single-node memory bottleneck and enables reconstructions at scales that were previously unattainable. (II) A new intelligent initialization module that generates a high-quality initial density field proposal via a fast approximate inversion. By doing so, it directly mitigates and substantially shortens the usually prolonged burn-in phase of the HMC sampler, yielding order-of-magnitude savings during burn-in. Consequently, the total computational cost is reduced by a substantial factor, accelerating convergence to a high-quality solution by several times.

The remainder of the paper is organized as follows. Section \ref{sec:2} describes our methodology, including the HMC formalism, the parallelization approach, and the new proposal-generation module. Section \ref{sec:3} provides an extensive performance evaluation, including strong and weak scaling, memory usage, and validation of reconstruction fidelity relative to simulations. In Section \ref{sec:discussion_conclusion}, we discuss the broader implications of our results and outline prospective avenues for future work. 


\section{An MPI code for Elucid-DESI}
\label{sec:2}

This work implements and extends the Bayesian reconstruction framework to obtain the initial condition of the structure formation of the local Universe based on the density field extracted from observation using the Hamiltonian Markov Chain Monte Carlo (HMCMC) method, developed by \citet{Wang_2013, Wang_2014}. The core idea is to sample the posterior distribution of the initial density field, $\delta_i$, conditioned on the observed final density field, $\delta_f$, by constructing a Hamiltonian system in which the potential energy is defined as the negative log-posterior.

We direct the readers to \citet{Wang_2013,Wang_2014} for a complete derivation of the Hamiltonian equations and the sampling algorithm. The focus of this section is to present our key computational advancements in this framework. Specifically, we will detail the following:

1. The overall reconstruction pipeline, highlighting the integration of the forward model and the sampler.

2. The theoretical Basis, introducing the Hamiltonian Markov Chain Monte Carlo (HMCMC) method.

3. The particle-mesh (PM) N-body solver, which serves as the forward model to evolve the density field and compute the likelihood gradient—the primary computational kernel.

4. A novel initial density field guess module, designed to significantly reduce the Markov chain's burn-in period.

These components form the foundation of our new, scalable MPI-parallel reconstruction code.

\begin{figure*}
	\includegraphics[width=1.8\columnwidth]{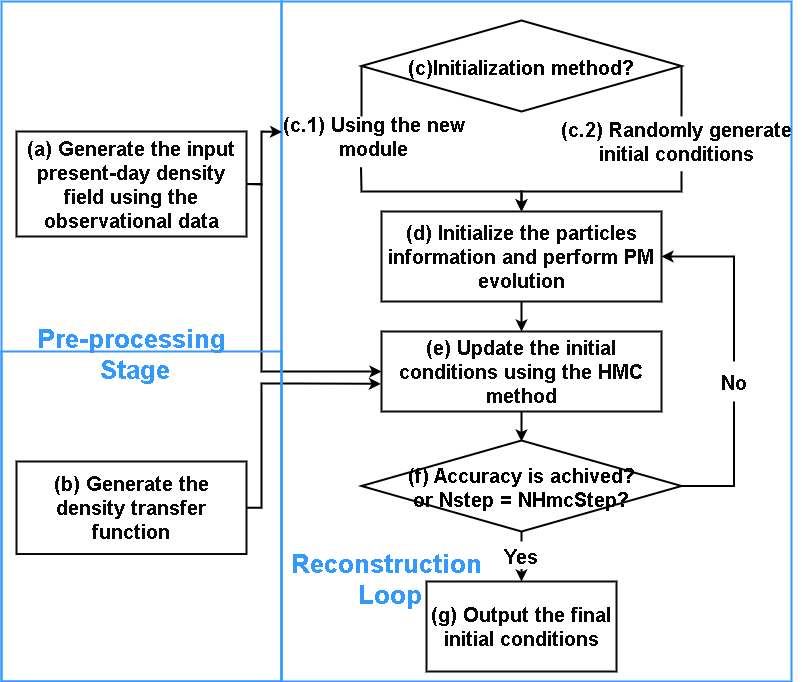}
    \caption{Schematic overview of the reconstruction pipeline. The workflow consists of two main stages. \textbf{Left panel:} Pre-processing stage, which prepares the necessary inputs: \textbf{(a)} reconstruction of the three-dimensional observed density field from survey data, and \textbf{(b)} calibration of the transfer function \(T(k)\) to correct for inaccuracies in the fast particle-mesh forward model.  \textbf{Right panel:} The core HMCMC reconstruction loop, which iteratively samples the initial density field. The process begins by \textbf{(c)} generating an initial condition guess—either \textbf{(c.1)} randomly or \textbf{(c.2)} via our novel \textit{guess module}—then \textbf{(d)} evolves it forward in time using the particle-mesh solver. Then, \textbf{(e)} the code evaluates the likelihood by comparing the predicted final field with the observed input, and updates the initial conditions via Hamiltonian dynamics. \textbf{(f)} This loop repeats until a predefined number of samples are collected or a required reconstruction accuracy is achieved. \textbf{(g)} The final reconstructed initial density field is outputted for subsequent studies. }
    \label{fig:flowchart}
\end{figure*}

\subsection{The flowchart}

Our new reconstruction code is built upon the scalable FastPM \citep{Feng_2016} scheme, a highly efficient approximate N-body solver. We have extensively modified and extended this framework to implement the Hamiltonian Markov Chain Monte Carlo (HMCMC) sampler required for Bayesian reconstruction of the initial density field.

The general workflow of the reconstruction pipeline is illustrated in Figure \ref{fig:flowchart}. It can be logically divided into two major phases: (1) a pre-processing stage (the two blue boxes on the left side of the figure) and (2) the core HMCMC sampling stage (the blue box on the right side).

\subsubsection{Pre-processing Stage: Construction of Inputs}

This stage prepares two essential inputs for the reconstruction code.

\begin{itemize}
    \item  \textbf{The Present-Day Density Field, \(\rho_{\text{obs}}\) (Figure~\ref{fig:flowchart}(a)):}
\end{itemize}

The target final density field shall be reconstructed from observational galaxy survey data. As this study mainly focuses on developing a highly efficient MPI code for the initial condition reconstruction, we will directly make use of the density field from a given N-body simulation. Observationally, we are constructing the mass, tidal tensor and velocity field from the DESI BGS data in a separate work (Li et al. 2026, in preparation). The procedures for obtaining the density field are described as follows:

\begin{enumerate}
    \item Identify galaxy groups from the survey data to serve as tracers of dark matter halos (e.g., using the group finder described in \citet{Yang05,Yang_2007}).
    \item Apply a mass threshold to select a robust halo sample.
    \item Reconstruct the continuous three-dimensional matter density field from the spatial distribution of these halos (e.g. following the techniques validated in \citet{Wang_2009,Wang12}).
    \item Based on the matter density field, we obtain the velocity field using linear theory.
    \item Move groups/halos to the real space according to the velocity field.
    \item Iteratively obtain the final matter density field.
\end{enumerate}
The resulting density field, \(\rho_{\text{obs}}\), serves as the constraining data \(\delta_f\) for the HMCMC likelihood.

\begin{itemize}
    \item \textbf{The Density Transfer Function, T(k) (Figure~\ref{fig:flowchart}(b)):} 
\end{itemize}

To bridge the accuracy gap between the FastPM approximation and a full 
N-body simulation, we employ a transfer function T(k). It is computed from a pair of simulations (one based on PM evolution, one based on standard simulation) that share the same initial conditions:
\begin{equation}
    T(k)=\frac{\langle\rho_{N,c}(\textbf{k})\rho_{si}^{\ast}(\textbf{k})\rangle_{\textbf{k}}}{\langle\rho_{N,c}(\textbf{k})\rho_{N,c}^{\ast}(\textbf{k})\rangle_{\textbf{k}}}.
	\label{eq:CorrectionCurve}
\end{equation}
Here, $\rho_{N,c}$ is the final density field obtained by cloud-in-cell (CIC) assignment after $N_{\text{PM}}$ FastPM steps (using the code's internal particle-mesh solver), and $\rho_{si}$  is the corresponding density field at $z$ = 0 from a high-accuracy reference simulation. This correction function is applied to the model-predicted density $\rho_{mod}$ field during sampling:
\begin{equation}
    \rho_{mod}(\textbf{k}) = \frac{w_g(R_sk)T(k)}{w_c(\textbf{k})N_c^3}\sum_x\rho_{N,c}(\textbf{x})e^{-i\textbf{k}\cdot\textbf{x}}
\label{eq:corrected_density}
\end{equation}
where $w_c$ is the CIC assignment kernel, $w_g(R_sk)$ is a Gaussian smoothing kernel with scale $R_s$ (introduced to suppress shot noise in the Hamiltonian force calculation), and $N_c$ denotes the number of grid cells in one dimension. 

These pre-processed inputs—the transfer function $T(k)$ and the target density field—are then supplied to the main reconstruction loop, which carries out the HMCMC sampling to infer the posterior distribution of the initial conditions $\delta_i$.

\subsubsection{Core HMCMC Sampling Stage: Reconstruction Loop}

The blue box on the right side of Figure \ref{fig:flowchart} highlights the main reconstruction loop, which carries out iterative Bayesian sampling using the HMCMC algorithm. The procedure unfolds as follows:

\begin{itemize}
\item \textbf{Initial Condition Generation(Figure~\ref{fig:flowchart}(c)):} 
\end{itemize}

The reconstruction begins by generating the initial guess $\delta_i$ for the initial density field. Our implementation provides two distinct initialization strategies:

\begin{enumerate}
\item \textbf{Conventional Random Initialization (Figure~\ref{fig:flowchart}(c.1)):} Following standard practice, the chain starts from a random Gaussian realization consistent with the theoretical power spectrum $P(k)$.

\item \textbf{Intelligent Initial Guess (Novel) (Figure~\ref{fig:flowchart}(c.2)):} Utilizing our newly developed \textit{initial guess module}, which generates a physically-informed initial proposal through fast, approximate reconstruction of the input density field $\rho_{\rm obs}$. This approach significantly accelerates convergence by initializing the chain near the high-probability region of the posterior distribution.
\end{enumerate}

\begin{itemize}
\item \textbf{Forward Model Evaluation(Figure~\ref{fig:flowchart}(d)):} 
\end{itemize}

The proposed initial field $\delta_i$ is evolved to the target redshift (typically $z=0$) using the integrated particle-mesh $N$-body solver. This produces a predicted final density field $\rho_{mod}$, which incorporates both gravitational evolution and the pre-computed correction transfer function $T(k)$ to enhance accuracy.

\begin{itemize}
\item \textbf{State Update and Likelihood Computation (Figure~\ref{fig:flowchart}(e)):} 
\end{itemize}

Following the evaluation, new proposals for both the density field and momentum variables are updated via Hamiltonian dynamics:
\begin{equation}
    \frac{d\delta_i}{dt} = \frac{\partial K}{\partial p}, \quad \frac{dp}{dt} = -\frac{\partial \psi}{\partial \delta_i}, \label{eq:moment}
\end{equation}
with numerical integration performed using a symplectic leapfrog scheme. Here, $K$ denotes the kinetic energy of the auxiliary momentum variables and $\psi$ represents the negative log-posterior (potential energy). This process returns to {\bf Forward Model Evaluation} step and continues for $N_{\mathrm{substeps}}$ iterations.

The final predicted field $\rho_{mod}$ is compared with the input density field $\rho_{p}$ to evaluate the Hamiltonian $\mathcal{H}$. The Metropolis-Hastings criterion determines whether to accept or reject the proposed state based on Hamiltonian conservation.


\begin{itemize}
\item \textbf{Chain Termination and Final Output(Figure~\ref{fig:flowchart}(f,g))} 
\end{itemize}

Upon completion of the pre-specified number of iterations $N_{\mathrm{steps}}$ or when the desired reconstruction accuracy is achieved, the Markov chain is terminated. In a full Bayesian analysis, one would retain all samples after burn-in to characterize the posterior distribution. For computational and storage efficiency in these large-scale runs, we output a single high-fidelity sample: the final accepted state $\delta_i^{\mathrm{(final)}}$ of the chain. 

We justify this choice by noting that a well-converged HMCMC chain spends the vast majority of its time in the high-probability region of the posterior. Therefore, the final state $\delta_i^{\mathrm{(final)}}$ is effectively a sample drawn from the stationary posterior distribution and serves as a valid, high-likelihood reconstruction. It can be considered an approximation to a sample from the posterior, suitable for applications such as generating constrained simulations. For uncertainty quantification, the full chain would be required.

Optimal reconstruction efficiency requires careful tuning of several key parameters within the HMCMC framework. These include the mass assignment scheme smoothing scale $R_s$, the time-stepping parameters in the leapfrog integrator, and the PM force resolution. While comprehensive guidance on parameter selection is beyond the scope of this paper, users may refer to the detailed discussions and tests presented in the original ELUCID implementation papers \citep{Wang_2013,Wang_2014} for established guidelines. Our code provides sensible defaults based on these references, but specific survey characteristics and scientific requirements may necessitate further optimization by the user.

\subsection{Theoretical Basis: Hamiltonian Monte Carlo for Density Field Reconstruction}
\label{sec:theory}

The goal of initial density field reconstruction is to find the initial fluctuation density field \(\delta(\mathbf{k})\) that, when evolved forward through a cosmological model, best matches an observed present-day density field $\rho_{\text{obs}}(\mathbf{x})$. This is formulated as a Bayesian inference problem with two fundamental constraints.

\subsubsection{Bayesian Formulation}

First, in accordance with the standard cosmological model, the initial density field is assumed to be a Gaussian random field characterized by the linear power spectrum \(P_{\text{lin}}(k)\). Its prior probability distribution in Fourier space is:
\begin{equation}
    P[\delta(\mathbf{k})] = \prod_{\mathbf{k}}^{\mathrm{half}} \prod_{j=0}^{1} 
    \frac{ \exp\left\{ -[\delta_j(\mathbf{k})]^{2} / P_{\text{lin}}(k) \right\} }
    { \left[ \pi P_{\text{lin}}(k) \right]^{1/2} },
    \label{eq:prior}
\end{equation}
where the product over \(\mathbf{k}\) extends over half of Fourier space (exploiting Hermitian symmetry), and \(j=0,1\) index the real and imaginary components, respectively. $P_{\text{lin}}(k)$ is the analytical linear power spectrum assuming a given set of cosmological parameters.

Second, the forward-evolved field \(\rho_{\text{mod}}(\mathbf{x})\), obtained from \(\delta(\mathbf{k})\) through a prescribed gravity solver, must match the input field (observed) \(\rho_{\text{obs}}(\mathbf{x})\). We quantify the mismatch using a weighted statistic \(\chi^2_{\omega}\),
\begin{equation}
    \chi^2_{\omega} = \sum_{\mathbf{x}} 
    \frac{ \left[ \rho_{\text{mod}}(\mathbf{x}) - \rho_{\text{obs}}(\mathbf{x}) \right]^2 \, \omega(\mathbf{x}) }
    { 2 \sigma_{\text{obs}}^2(\mathbf{x}) },
    \label{func:chi2}
\end{equation}
where \(\sigma_{\rm obs}(\mathbf{x})\) represents the observational uncertainty and \(\omega(\mathbf{x})\) is a survey window function.

Assuming a Gaussian likelihood \(\mathcal{L} \propto \exp(-\chi^2_{\omega})\), the posterior distribution for the initial field given the data is:
\begin{equation}
\begin{aligned}
    Q(\delta(\textbf{k})|\rho_{\text{obs}}(\textbf{x})) &= e^{-\chi^2_{\omega}}\times P[\delta(\textbf{k})] \\
    & = e^{-\sum_{\textbf{x}} {[\rho_{\text{mod}}(\textbf{x})-\rho_{\text{obs}}(\textbf{x})]^2\omega(\textbf{x})}/{2\sigma_{\rm obs}^2(\textbf{x})}}\\
    &\quad\times \prod_{\textbf{k}}^{\text{half}}\prod_{j=0}^1\frac{\exp\{-[\delta_j(\textbf{k})]^{2}/P_{\text{lin}}(k)\}}{[\pi P_{\text{lin}}(\text{k})]^{1/2}}\,.
\end{aligned}
\end{equation}
The reconstruction problem thus reduces to sampling from this high-dimensional posterior to find the most probable initial configuration.

\subsubsection{Hamiltonian Monte Carlo Implementation}

We employ Hamiltonian Monte Carlo (HMC) to efficiently sample the posterior
distribution \(Q\). First, we define the "potential energy" function as the negative log-posterior:
\begin{equation}
    \begin{aligned}
        \psi[\delta(\textbf{k})] &\equiv -\ln Q[\delta_j(\textbf{k}) \, | \, \rho_{\rm obs}(\textbf{x})] \\
        &= \sum_{\mathbf{k}}^{\mathrm{half}} \ln\!\big[ \pi P_{\text{lin}}(k) \big]
           + \sum_{\mathbf{k}}^{\mathrm{half}} \sum_{j=0}^{1} \frac{[\delta_j(\mathbf{k})]^2}{P_{\text{lin}}(k)} \\
        &\quad + \sum_{\mathbf{x}} 
           \frac{ \left[ \rho_{\text{mod}}(\mathbf{x}) - \rho_{\rm obs}(\mathbf{x}) \right]^2 \omega(\mathbf{x}) }
                { 2 \sigma_{\text{obs}}^2(\mathbf{x}) }.
    \end{aligned}
    \label{eq:potential}
\end{equation}
Building on this framework, we introduce conjugate momentum variables \(p_j(\mathbf{k})\) and auxiliary "mass" parameters \(m_j(\mathbf{k})\), and define the Hamiltonian as
\begin{equation}
    H[\delta, p] = 
    \sum_{\mathbf{k}}^{\mathrm{half}} \sum_{j=0}^{1} \frac{p_j^2(\mathbf{k})}{2 m_j(\mathbf{k})} 
    + \psi[\delta(\textbf{k})].
    \label{eq:hamiltonian}
\end{equation}

The associated partition function can be factorized as
\begin{equation}
    \exp(-H) = Q[\delta(\textbf{k}) \, | \, \rho_{\text{obs}}(\textbf{x})] \; 
    \prod_{\mathbf{k}}^{\mathrm{half}} \prod_{j=0}^{1} 
    \exp\!\left[ -\frac{p_j^2(\mathbf{k})}{2 m_j(\mathbf{k})} \right],
    \label{eq:partition}
\end{equation}
which shows that drawing samples from the joint distribution \(\exp(-H)\) and integrating out the momenta produces samples from the desired posterior \(Q\).

Following the prescription of \citet{Wang_2013,Wang_2014}, we define the Hamiltonian mass as:

\begin{equation}
    m_j(k) \equiv m(k) = \frac{2}{P_{\mathrm{lin}}(k)} + \sqrt{ \frac{ \left\langle \sum_{j=\mathrm{re,im}} F_j^2(\mathbf{k}) \right\rangle_k }{ P_{\mathrm{lin}}(k) } },
    \label{eq:mass_definition}
\end{equation}

where $\langle \cdots \rangle_k$ denotes an average over the phase of $k$, and ${F}_j  (\mathbf{k}) \equiv \frac{\partial \chi^2}{\partial\delta_j{\textbf{(k)}}}$ represents the Fourier-space force components. This definition is motivated by the analysis in \citet{Wang_2013} and improves the sampling efficiency of the HMC chain.

At the beginning of each HMC step, we draw the conjugate momenta $p_j(\mathbf{k})$ independently from a Gaussian distribution with zero mean and variance $m_j(k)$:
\begin{equation}
    p_j(\mathbf{k}) \sim \mathcal{N}\big(0,\, m_j(k)\big).
    \label{eq:momentum_sampling}
\end{equation}
This choice ensures that the kinetic energy term $\sum p_j^2/(2m_j)$ follows a chi-squared distribution with the correct scale, leading to efficient exploration of the posterior.

\subsubsection{Sampling Algorithm}

The HMC algorithm proceeds by simulating Hamiltonian dynamics in phase space \((\delta, p)\) with
\begin{equation}
    \frac{d\delta_j(\mathbf{k})}{dt} = \frac{\partial H}{\partial p_j(\mathbf{k})}, \qquad
    \frac{dp_j(\mathbf{k})}{dt} = -\frac{\partial H}{\partial \delta_j(\mathbf{k})}.
    \label{eq:hamilton_eqs}
\end{equation}
In practice, these equations are integrated numerically using a symplectic leapfrog scheme. Starting from a current state \([\delta_j(\textbf{k}), p_j(\textbf{k})]\), a trajectory of length \(\tau\) (defined by the number of leapfrog steps and step size) yields a proposed new state \([\delta'_j(\textbf{k}), p'_j(\textbf{k})]\). The proposal is accepted with probability:
\begin{equation}
    P_{\text{acpt}} = \min\!\left\{ 1, \; 
    \exp\!\big( -[H(\delta'_j(\textbf{k}), p'_j(\textbf{k})) - H(\delta_j(\textbf{k}), p_j(\textbf{k}))] \big) \right\}.
    \label{eq:acceptance}
\end{equation}
By repeating this "Hamiltonian trajectory + Metropolis update" cycle, the Markov chain explores the posterior distribution. After iterating for a sufficiently large number of steps \(N_{\mathrm{steps}}\), once the \(\chi^2_{\omega}\) value has converged, the resulting initial density field \(\delta_j(\textbf{k})\) is taken as the final output.

\subsubsection{Likelihood Gradient via Adjoint Differentiation}
\label{sec:likelihood_gradient}

The likelihood term of the Hamiltonian force, denoted as ${F}_j(\mathbf{k})$, is derived from the derivative of the $\chi^2$ statistic with respect to the initial density field $\delta_j(\mathbf{k})$. Following the chain rule, this gradient can be expressed as
\begin{equation}
{F}_j(\mathbf{k}) = \frac{\partial \chi^2}{\partial \delta_j(\mathbf{k})}
= \frac{\partial \chi^2}{\partial \mathbf{p}_N} \otimes
\frac{\partial \mathbf{p}_N}{\partial \mathbf{p}_{N-1}} \otimes \cdots \otimes
\frac{\partial \mathbf{p}_1}{\partial \mathbf{p}_0} \otimes
\frac{\partial \mathbf{p}_0}{\partial \delta_j(\mathbf{k})},
\label{eq:gradient_chain}
\end{equation}
where $\mathbf{p}_n = [\mathbf{r}_n, \mathbf{v}_{n-1/2}]$ denotes the particle positions and velocities at the $n$-th PM step, and $\otimes$ represents matrix multiplication.

Directly computing this product in the forward direction would involve prohibitively large intermediate matrices. Instead, we adopt the \textit{adjoint differentiation} technique \citep{Hanson_Cunningham_1996}, which evaluates the gradient by reversing the order of multiplication:

\begin{equation}
\frac{\partial \chi^2}{\partial \delta_j(\mathbf{k})} =
\left( \cdots \left( \left( \frac{\partial \chi^2}{\partial \mathbf{p}_N} \cdot
\frac{\partial \mathbf{p}_N}{\partial \mathbf{p}_{N-1}} \right) \cdot
\frac{\partial \mathbf{p}_{N-1}}{\partial \mathbf{p}_{N-2}} \right) \cdots \right) \cdot
\frac{\partial \mathbf{p}_0}{\partial \delta_j(\mathbf{k})}.
\label{eq:adjoint_product}
\end{equation}

In practice, this reverse-mode differentiation proceeds in three stages, mirroring the forward PM evolution:

\begin{enumerate}
    \item \textbf{$\chi^2$ transformation:} Compute the derivative of $\chi^2$ with respect to the final particle positions $\partial \chi^2 / \partial \mathbf{r}_N$. The velocity derivative $\partial \chi^2 / \partial \mathbf{v}_{N-1/2}$ is set to zero, as $\chi^2$ depends only on the density field derived from the final particle positions.
    
    \item \textbf{PM back-propagation:} Working backwards through the $N$ PM steps, propagate the derivatives from $\mathbf{p}_{n+1}$ to $\mathbf{p}_n$ using the adjoint of the PM update equations. This step provides $\partial \chi^2 / \partial \mathbf{r}_0$ and $\partial \chi^2 / \partial \mathbf{v}_0$ at the initial time.
    
    \item \textbf{Zel'dovich back-propagation:} Finally, relate the initial particle derivatives to the derivative with respect to the initial density field $\delta_j(\mathbf{k})$ via the Zel'dovich approximation. This yields the likelihood term ${F}_j(\mathbf{k})$ of the Hamiltonian force.
\end{enumerate}

The detailed derivation of each adjoint step follows the same principles as described in \citet{Wang_2014} (see their Appendix B), but has been consistently adapted to our FastPM-based forward model \citep{Feng_2016}. As a result, the entire gradient calculation requires only a single reverse pass through the PM time steps, with a computational cost comparable to one forward simulation.

\subsection{Particle-Mesh Gravity Solvers}
\label{sec:pm_method}

Our reconstruction code integrates two distinct particle-mesh (PM) $N$-body evolution schemes as forward models within the HMCMC sampler. The choice between them allows users to balance computational speed against required accuracy for specific applications.

Both integration schemes described below operate in a comoving cosmological framework. We define the dimensionless Hubble parameter \(E(a) \equiv H(a)/H_0\), where \(H(a) = H_0 \sqrt{\Omega_{m,0} a^{-3} + \Omega_{\Lambda,0}}\) for a flat \(\Lambda\)CDM cosmology. The scale factor \(a\) evolves from \(a_{\mathrm{ini}} = 1/(1+z_{\mathrm{ini}})\) to \(a_{\mathrm{final}} = 1\). The derivative \(\dot{a} = da/dt = a H(a) = a H_0 E(a)\) is used in the time integration factors. Importantly, both the KDK and FastPM integrators are symplectic, a property that ensures long-term conservation of the Hamiltonian — a critical requirement for stable and accurate HMC sampling.

\subsubsection{Traditional Kick-Drift-Kick (KDK) Scheme}

The first option is a conventional second-order kick-drift-kick (KDK) leapfrog integrator, which advances particle positions $\mathbf{r}$ and velocities $\mathbf{v}$ (in comoving coordinates) via:
\begin{align}
    \mathbf{v}(\mathbf{q}, a_{n+1/2}) &= \mathbf{v}(\mathbf{q}, a_{n-1/2}) 
    - \nabla\Phi(\mathbf{r}(\mathbf{q}, a_n)) \int_{a_{n-1/2}}^{a_{n+1/2}} 
    \frac{4\pi G\bar{\rho}_0}{a\dot{a}} \, da , \label{eq:kd_kick} \\
    \mathbf{r}(\mathbf{q}, a_{n+1}) &= \mathbf{r}(\mathbf{q}, a_n) 
    + \mathbf{v}(\mathbf{q}, a_{n+1/2}) \int_{a_n}^{a_{n+1}} 
    \frac{1}{a^2\dot{a}} \, da, \label{eq:kd_drift}
\end{align}
where $\mathbf{q}$ denotes the Lagrangian coordinate, $\Phi$ is the gravitational potential computed on the mesh, $\bar{\rho}_0$ is the mean matter density today, and $a$ is the scale factor with $\dot{a} = da/dt$. $n$ is the step number and $a_{n+1/2} = (a_n + a_{n+1})/2$.

For the sake of clarity, we rewrite these equations in a form that separates the cosmology-dependent integration factors from the instantaneous dynamical terms. The dimensionless integration factors for a step from \(a_0\) to \(a_1\) are defined as:
\begin{align}
    \mathcal{D}_{\mathrm{pm}}(a_0, a_1) &\equiv \int_{a_0}^{a_1} \frac{1}{a^3 E(a)}  \, da, \label{eq:dpm_def} \\
    \mathcal{K}_{\mathrm{pm}}(a_0, a_1) &\equiv \int_{a_0}^{a_1} \frac{1}{a^2 E(a)}  \, da. \label{eq:kpm_def}
\end{align}
The KDK update equations can then be expressed compactly as:
\begin{align}
    \mathbf{v}(\mathbf{q}, a_{n+1/2}) &= \mathbf{v}(\mathbf{q}, a_{n-1/2}) 
    + \mathbf{F}_n(\mathbf{r}_n(\mathbf{q})) \; \mathcal{K}_{\mathrm{pm}}(a_{n-1/2}, a_{n+1/2}), \label{eq:kd_kick_compact} \\
    \mathbf{r}(\mathbf{q}, a_{n+1}) &= \mathbf{r}(\mathbf{q}, a_n) 
    + \mathbf{P}_{n+1/2}(\mathbf{v}_{n+1/2}(\mathbf{q})) \; \mathcal{D}_{\mathrm{pm}}(a_n, a_{n+1}), \label{eq:kd_drift_compact}
\end{align}
where
\begin{align}
    \mathbf{F}_n(\mathbf{r}_n(\mathbf{q})) &\equiv -\nabla\Phi(\mathbf{r}_n(\mathbf{q})) \times \frac{4\pi G\bar{\rho}_0}{H_0}, \\
    \mathbf{P}_{n}(\mathbf{v}_{n}(\mathbf{q})) &\equiv \mathbf{v}_{n}(\mathbf{q}) \times H_0.
\end{align}
Here, \(\mathbf{F}_n\) is the scaled gravitational force and \(\mathbf{P}_{n+1/2}\) is the scaled momentum. The prefactors \(4\pi G\bar{\rho}_0/H_0\) and \(H_0\) are absorbed into these definitions to make the factors \(\mathcal{D}_{\mathrm{pm}}\) and \(\mathcal{K}_{\mathrm{pm}}\) purely geometric integrals, as defined in Eqs.~\eqref{eq:dpm_def} and \eqref{eq:kpm_def}.

This scheme assumes constant force and velocity within each time step, performing alternating ``kick'' (velocity update) and ``drift'' (position update) operations.

Although it is robust and widely adopted, the KDK scheme produces larger amplitude and phase errors on large scales when using coarse time stepping, because it does not perfectly reproduce the linear growth of cosmological perturbations. Consequently, achieving high accuracy demands a comparatively large number of time steps, which raises the computational cost.

\subsubsection{FastPM Modified-Factor Scheme}

The second option implements the \textit{FastPM} algorithm developed by \citet{Feng_2016}, which modifies the kick and drift factors to ensure exact linear growth on large scales even with finite time steps. This scheme is derived from the equations of motion in the Zeldovich approximation and corrects the integration factors to match linear theory evolution exactly for the displacement and velocity fields.

The modified drift ($\mathcal{D}_{\mathrm{FASTPM}}$) and kick ($\mathcal{K}_{\mathrm{FASTPM}}$) factors are defined as:
\begin{align}
    \mathcal{D}_{\mathrm{FASTPM}} &= \frac{ \Delta[x_{\mathrm{ZA}}]_{a_0}^{a_1} }{ p_{\mathrm{ZA}} }
    = \frac{1}{a_r^3 E(a_r)} \left( \frac{ \Delta[G_p]_{a_0}^{a_1} }{ g_p(a_r) } \right), \label{eq:fastpm_drift} \\
    \mathcal{K}_{\mathrm{FASTPM}} &= \frac{ \Delta[p_{\mathrm{ZA}}]_{a_0}^{a_1} }{ f_{\mathrm{ZA}} }
    = \frac{1}{a_r^2 E(a_r)} \left( \frac{ \Delta[G_f]_{a_0}^{a_1} }{ g_f(a_r) } \right), \label{eq:fastpm_kick}
\end{align}
where $a_r$ is a reference scale factor (typically the midpoint of the interval) and $D(a)$ is the linear growth factor. The auxiliary functions are:
\begin{align}
    g_p(a) &= \frac{dD}{da}, \label{eq:gp} \\
    g_f(a) &= \frac{d}{da}\left[ a^3 E(a) g_p(a) \right]
    = 3a^2 E \frac{dD}{da} + a^3 \frac{dE}{da} \frac{dD}{da} + a^3 E \frac{d^2D}{da^2}, \label{eq:gf} \\
    G_p(a) &= D(a), \quad G_f(a) = a^3 E(a) g_p(a). \label{eq:G_functions}
\end{align}
These correction factors explicitly account for the evolution of the cosmological scale factor, guaranteeing that displacements, velocities, and forces follow the behavior predicted by linear perturbation theory on large scales, independent of the chosen time step size.

In the limit of infinitesimal time steps, the FastPM factors reduce to the standard KDK factors defined in Eqs.~\eqref{eq:dpm_def} and \eqref{eq:kpm_def}.

\subsubsection{Comparison and Usage}

The FastPM and KDK integrators differ in both per-step cost and the number of steps required to reach a given accuracy. The traditional KDK scheme has a lower cost per step, following a sequence of three operations: force evaluation (F), velocity update (V), and position drift (P), commonly denoted as FVP. In contrast, the FastPM scheme involves approximately five operations per step — velocity kick, position drift, force evaluation, and two modified-factor corrections (VPPFV) — making it more expensive on a per-step basis. However, FastPM achieves the same level of accuracy as KDK with significantly fewer steps. Therefore, for high-precision reconstructions such as our ELUCID-DESI pipeline, the reduced total step count of FastPM outweighs its higher per-step cost, making it the computationally preferred choice. For very high-resolution simulations where small-scale forces dominate, the traditional KDK scheme remains a viable option, though at the cost of increased time steps to maintain comparable large-scale accuracy.

\subsection{Accelerating Convergence: An Initial Density Field Guess Module}
\label{sec:guess_module}

One of the main computational burdens of HMCMC-based reconstruction stems from the lengthy “burn-in” period required for the Markov Chain to transition from an arbitrary initial state into the high-probability region of the posterior distribution. To mitigate this, we introduce a rapid, approximate method that generates an informed initial estimate of the density field. This estimate serves as a high-quality starting point for the HMCMC sampler and significantly cuts down the number of steps needed to achieve convergence.

\subsubsection{Motivation and Application}

As shown in the right panel of Figure \ref{fig:flowchart}, the standard method begins the chain from a random Gaussian field sampled from the prior power spectrum $P_{\mathrm{lin}}(k)$. While this initialization is statistically valid, it typically lies far from the desired posterior mode, so the chain often requires hundreds to thousands of costly forward simulations before it reaches an efficient sampling regime.

Our new module introduces an alternative route for initialization: it quickly generates a physically plausible estimate $\delta_{\mathrm{guess}}(\mathbf{k})$ that already closely matches the true initial conditions. This estimate is then adopted as the initial state $\delta_i^{(0)}$ for the HMCMC chain, thereby skipping most of the burn-in period. The module runs entirely as a pre-processing step and adds only negligible overhead to the overall pipeline.

\subsubsection{Algorithm: Approximate Linear Inversion}

The guess of the initial density field is generated through an approximate linear inversion of the input final density field $\rho_{\text{p}}(\mathbf{x})$, inspired by the same transfer-function correction used in the forward model (Section 2.1). The procedure is as follows:

\begin{itemize}
    \item  \textbf{Compute the relative change kernel} 
\end{itemize}

Using simulated data at both an initial and a final redshift, we define a scale-dependent amplitude transfer function in Fourier space:
\begin{equation}
    T_{\mathrm{guess}}(k) = \sqrt{ \frac{D(z)^2P_{\mathrm{k}}(z=0, k)}{P_{\mathrm{k}}(z=z_{\mathrm{ini}}, k)} },
    \label{eq:guess_transfer}
\end{equation}
where $P_{\mathrm{k}}(z, k)$ represents the power spectrum measured from the reference simulation. In principle, a high-accuracy estimate of this function should be obtained from a more accurate reference simulation (e.g., using an $N$-body code such as \textsc{Gadget-4}, as opposed to the fast PM simulation). However, running an accuracy simulation solely for this kernel is computationally prohibitive, especially when preparing for large-scale reconstructions. To conserve substantial resources, we instead compute this function using the \textbf{ CSST Emulator}, which provides a fast and acceptably accurate approximation for the corresponding cosmology (see Appendix \ref{app:calibration_data}).

\begin{itemize}
    \item \textbf{Apply approximate inversion:}
\end{itemize}    

The input field $\rho_{\text{obs}}(\mathbf{x})$ is first smoothed with a kernel $R_{\mathrm{sml}}$ and corrected for the CIC assignment window $W_{\mathrm{cic}}$ using a mesh with the same number of grid cells as the density field. It is then inversely weighted by $T_{\mathrm{guess}}(k)$ to approximately "rewind" gravitational growth:
    \begin{equation}
        \delta_{\mathrm{guess}}(\mathbf{k}) = \frac{ \widetilde{\rho}_{\text{p}}(\mathbf{k}) }{ T_{\mathrm{guess}}(k)  \, R_{\mathrm{sml}}(k) \, W_{\mathrm{cic}}(k) }.
        \label{eq:guess_field}
    \end{equation}
Here $\widetilde{\rho}_{\text{obs}}(\mathbf{k})$ is the Fourier transform of the pre-processed observed field on grids. The product $ R_{\mathrm{sml}} \, W_{\mathrm{cic}}$ constitutes the combined kernels mentioned above.
    
\begin{itemize}
    \item \textbf{Output:} 
\end{itemize}   

The resulting field $\delta_{\mathrm{guess}}(\mathbf{k})$ is transformed back to real space and serves as the proposed guess of the initial density field.

\subsubsection{Advantages}

The guess field $\delta_{\mathrm{guess}}$ is \textit{not} intended as a standalone reconstruction. It neglects mode coupling, non-Gaussianities induced by non-linear evolution, and posterior uncertainties. Consequently, its accuracy is limited, particularly on small scales where non-linear effects are strong.

However, its primary advantage is that it delivers a \textit{physically informed initial estimate} that is already correlated with the true starting conditions. When used to initialize the HMCMC chain, $\delta_{\mathrm{guess}}$ generally falls within the attraction basin of the posterior mode, allowing the sampler to reach equilibrium orders of magnitude more quickly than when starting from a random state. As shown in Section \ref{sec:new_module_result}, this leads to substantial reductions in CPU time, particularly for large-volume reconstructions where burn-in would otherwise require several hundred forward simulation steps.

This module represents a classic trade-off between accuracy and efficiency. On one hand, the guess field neglects mode coupling, non-Gaussianities, and small-scale details, and therefore cannot be used as a final reconstruction. On the other hand, this approximation requires a small upfront investment in fast, low-accuracy preprocessing. In return, it yields disproportionately large gains in the efficiency of the subsequent high-fidelity (but expensive) Bayesian sampling, as demonstrated in Section~\ref{sec:new_module_result}. Thus, the module sacrifices a degree of final precision (which the HMCMC chain later restores) for a dramatic reduction in computational cost.

\section{Reconstruction Performance:}
\label{sec:3}

We evaluate the performance of our MPI-parallelized reconstruction code using a series of tests carried out on the Siyuan computing cluster. Our assessment targets three main aspects: (1) computational scalability with respect to both the reconstruction size and the number of processors, (2) memory usage efficiency, (3) the accuracy of the reconstructions, and (4) the speedup provided by the new initial guess module. 

\subsection{Simulations used to evaluate performance}
\label{sec:test_setup}

Since this study primarily aims to develop a highly efficient MPI code for initial-condition reconstruction, we directly utilize the density field from pre-existing N-body simulations. The final density fields used as inputs for the reconstruction are taken from cosmological $N$-body simulations, under the assumption of a spatially flat $\Lambda$CDM cosmology. To ensure robustness, we employ three independent sets of simulations:

\begin{itemize}
    \item \textbf{Gadget-4 Simulations} 
\end{itemize}    

For this work, we run a set of cosmological simulations using \textsc{Gadget-4} \citep{Gadget4} by ourselves, with the cosmological parameters $\Omega_{m,0}=0.3111$, $\Omega_{\Lambda,0}=0.6889$, $\Omega_{b,0}=0.0489$, $h=0.6766$, $n_s=0.9667$, and $\sigma_8=0.8159$. The initial linear power spectra are generated using \textsc{CLASS} \citep{Diego_Blas_2011}. The initial conditions for all simulations were generated at $z_{\mathrm{ini}} = 127$ using the Zeldovich approximation, which is the standard approach for the simulation codes used in this work.

\begin{itemize}
    \item \textbf{Kun Simulations}
\end{itemize}  

The second group of simulations comes from the Kun simulation project \citep{Chen_2025}, which is part of the Jiutian simulation suite \citep{han2025jiutiansimulationscsstextragalactic}. For consistency, the fiducial run adopts the same $\Lambda$CDM cosmological parameters as described above.

\begin{itemize}
    \item \textbf{Quijote Simulations} 
\end{itemize}  

We also utilize data from the \textsc{Quijote} simulation suite \citep{Quijote_sims} for low-resolution accuracy tests, ensuring that our results are not specific to a single simulation code.

To investigate performance on different scales, we use simulation boxes of size $L_{\mathrm{box}} = 1000\,h^{-1}\mathrm{Mpc}$ with varying particle counts: $256^3$, $512^3$, and $1024^3$. All simulations share a common redshift starting point of $z_{\mathrm{ini}} = 127$. The corresponding final density fields at $z=0$ serve as the observational targets $\rho_{\text{obs}}$ in our reconstruction tests, whereas the true initial conditions are stored for subsequent accuracy validation.


\begin{figure*}
	\includegraphics[width=1.8\columnwidth]{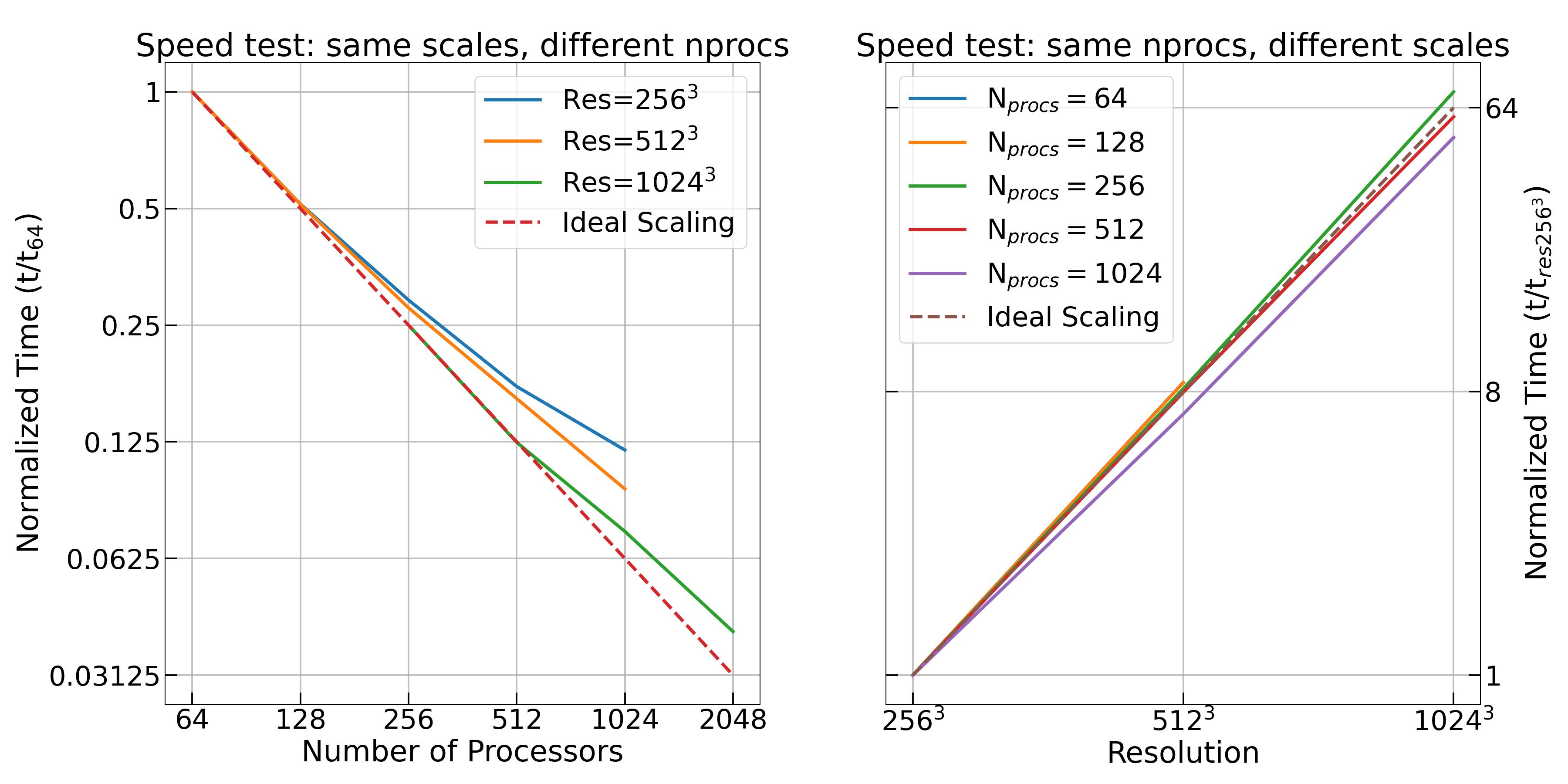}
     \caption{Parallel scaling performance. \textbf{(Left) Processor-number scaling} for fixed particle numbers ($256^3$, $512^3$, $1024^3$ particles). Runtime per iteration is normalized to show parallel efficiency relative to ideal linear scaling (dashed line). Larger problems scale better. \textbf{(Right) Particle-number scaling} at fixed MPI process counts. Runtime scales nearly linearly with total particle count (dashed line). The flatter slopes for higher process counts originate from load imbalance at the smallest problem size used for normalization (see Section~\ref{sec:speed}).}
    \label{fig:Speed}
\end{figure*}

\subsection{Scaling Performance}
\label{sec:speed}

To evaluate the parallel efficiency and computational complexity of our MPI implementation, we conduct two complementary types of scaling tests. 

\begin{itemize}
\item \textbf{Processor-number scaling} 
evaluates how the wall-clock time decreases as the number of MPI processes ($N_{\mathrm{proc}}$) increases, while the total number of particles (or grids) remains constant. This probes how efficiently the code parallelizes for a fixed problem size.
\end{itemize}

\begin{itemize}
\item \textbf{Particle-number scaling} (at fixed parallelism) 
evaluates how the wall-clock time grows with increasing total particle numbers when the number of MPI processes is kept fixed. This reflects the computational complexity of the algorithm for a given parallel setup and is essential to predict the cost of larger-scale simulations.
\end{itemize}

We conduct experiments with three problem sizes containing $256^3$, $512^3$, and $1024^3$ particles. For the $256^3$ and $512^3$ configurations, we employ $N_{\mathrm{proc}} = 64, 128, 256, 512,$ and $1024$. For the largest configuration, $1024^3$, we extend the scaling study to higher parallelism with $N_{\mathrm{proc}} = 256, 512, 1024,$ and $2048$. The lower bound of $N_{\mathrm{proc}}=256$ is imposed because simulations with fewer processes would require excessive runtime for repeated benchmarks. All reported timings correspond to the wall-clock time required for ten full HMCMC iterations (each consisting of one forward simulation and one Hamiltonian update). The outcomes are presented in Figure~\ref{fig:Speed}.

\subsubsection{Processor-number scaling}

The left panel of Figure~\ref{fig:Speed} presents the Processor-number scaling performance. To compare the scaling trends of different particle numbers on a unified plot, we normalize the runtime for each to a common reference point on the ideal scaling line (dashed). 

The $256^3$ and $512^3$ particle runs are normalized by their runtime at $N_{\mathrm{proc}}=64$, which serves as the anchor point (normalized time = 1 at $N_{\mathrm{proc}}=64$). 

For the $1024^3$ particle runs, we also wish to anchor them to the same ideal line starting at $N_{\mathrm{proc}}=64$. However, our smallest process count for the $1024^3$ case is $N_{\mathrm{proc}}=256$. Therefore, we first normalize the $1024^3$ particle runtime by its value at $N_{\mathrm{proc}}=256$. Then, to correctly place this starting point on the ideal line relative to the $N_{\mathrm{proc}}=64$ anchor, we apply a correction. Since increasing from 64 to 256 processes (a factor of 4) under ideal strong scaling would reduce the runtime by a factor of 4, we divide the normalized $1024^3$ time series by the same factor of 4. This operation effectively projects its performance trend onto the same reference ideal scaling line as the smaller problem sizes, allowing a direct visual comparison of how parallel efficiency degrades with increasing $N_{\mathrm{proc}}$ across different particle numbers.

All test cases exhibit a substantial speed-up as $N_{\mathrm{proc}}$ increases, even though the scaling efficiency steadily decreases. This departure from ideal scaling is anticipated and can be ascribed to two primary factors:
\begin{itemize}
    \item \textbf{Communication overhead:} As $N_{\mathrm{proc}}$ grows, the relative cost of MPI communication (for particle passing, force synchronization, and Fourier transforms) increases compared to local computation.
\end{itemize}
    
\begin{itemize}
    \item \textbf{Load imbalance:} Our domain decomposition divides the simulation volume into spatial subdomains. As the number of these subdomains grows, statistical fluctuations in the number of particles per subdomain become more significant, which in turn increases the imbalance of computational workload across processes.
\end{itemize}

However, as the number of particles increases, the loss in efficiency becomes smaller. The $512^3$ case exhibits better scaling efficiency—its curve stays nearer to the ideal line—than the $256^3$ case. This is because a larger problem size results in more computation per process for a fixed $N_{\mathrm{proc}}$, which more effectively spreads out the cost of the constant communication and synchronization overhead.

Importantly, this pattern persists for the even larger $1024^3$ case. Although its normalized values are shifted because of the different normalization reference (as discussed above), the slope of its curve—reflecting how quickly efficiency is lost as $N_{\mathrm{proc}}$ grows—is the most favorable of the three. Among all cases, the $1024^3$ curve deviates the least from ideal scaling as $N_{\mathrm{proc}}$ increases from 256 to 2048. This strongly indicates that our parallel implementation becomes progressively more efficient for the large problem sizes that are the main focus of this work, as the computation-to-communication ratio improves.

\begin{figure*}
    \centering
    \includegraphics[width=0.85\linewidth]{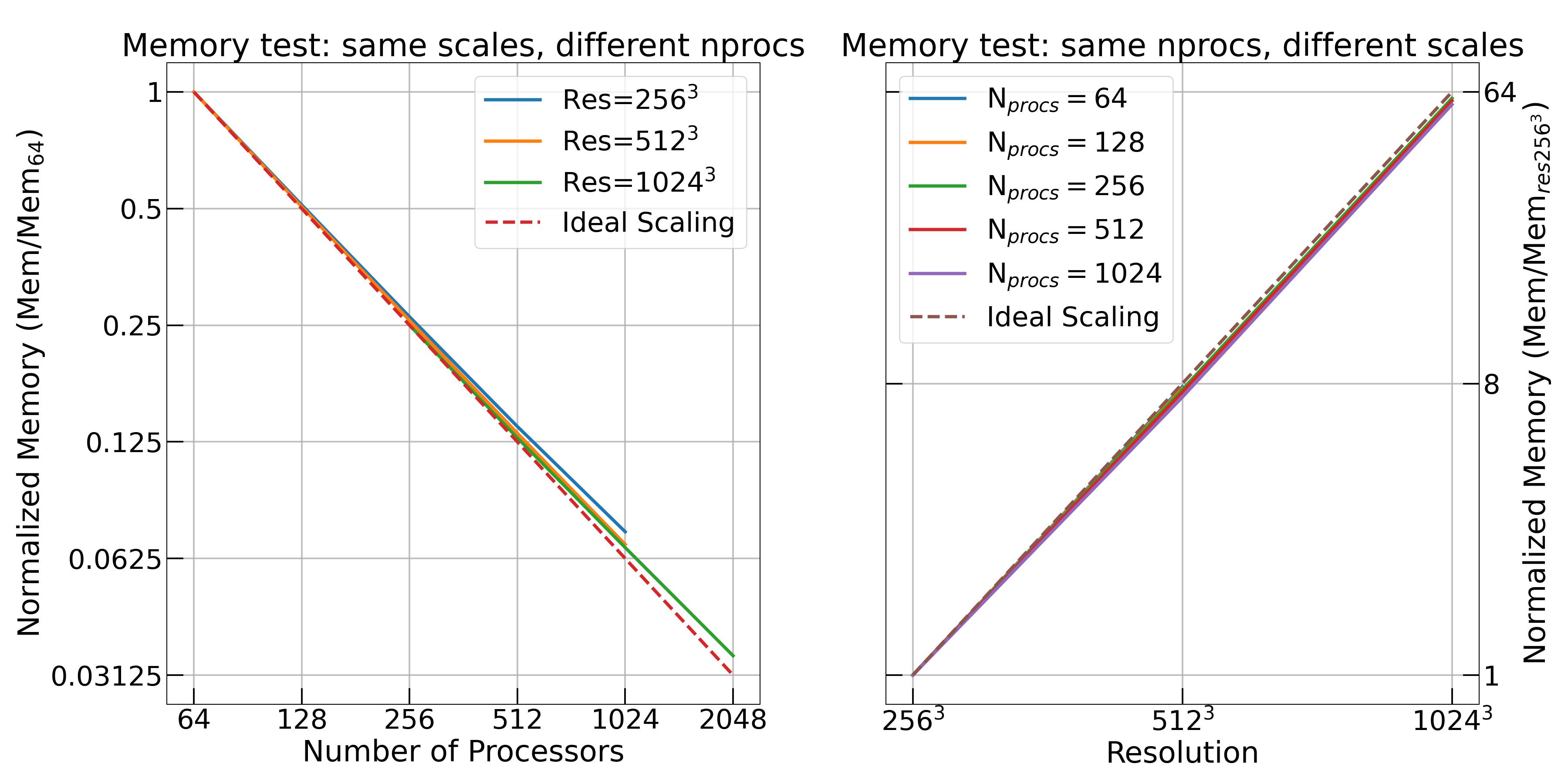}
    \caption{Memory usage per MPI process. \textbf{(Left) Processor-number scaling:} Memory per process for fixed reconstruction particle numbers ($256^3$, $512^3$, $1024^3$), normalized to show efficient reduction relative to ideal scaling (dashed line). \textbf{(Right) Particle-number scaling:} Memory per process as a function of total particle count, with the number of MPI processes held constant for each curve. All curves exhibit highly linear scaling, closely following the ideal dashed line, confirming predictable memory requirements essential for large-scale initial condition reconstruction (see Section~\ref{sec:memoryusage}).}
\label{fig:Memory}
\end{figure*}

\subsubsection{Particle-number scaling}

The right panel of Figure~\ref{fig:Speed} examines how computational cost grows with the total number of particles while keeping the parallel resources fixed at a given value of $N_{\mathrm{proc}}$. For each chosen process count (e.g., 64, 128, 256, 512), we record the wall-clock time per iteration as the total number of particles is increased from $256^3$ up to $1024^3$.

The results indicate that, for each fixed value of $N_{\mathrm{proc}}$, the runtime grows roughly linearly with the total number of particles, as evidenced by the curves largely tracking the ideal dashed line. Such linear behavior is anticipated because the main computational kernels exhibit a time complexity that is nearly $O(N)$ with respect to the particle or grid count over the range investigated.

An important pattern to note is that the slopes of these curves decrease as $N_{\mathrm{proc}}$ grows (for example, the blue curve for $N_{\mathrm{proc}}=64$ has a steeper slope than the purple curve for $N_{\mathrm{proc}}=1024$). This behavior arises directly from the load imbalance effect described in the Processor-number scaling analysis. 

Specifically, for the smallest problem size ($256^3$), using a large $N_{\mathrm{proc}}$ leads to significant load imbalance. Each process holds only a few particles, so fluctuations in particle counts across subdomains become large compared to the average. Consequently, the observed runtime for the $256^3$ case is longer than it would be with ideal load balance. Because this $256^3$ runtime is used as the normalization baseline for each $N_{\mathrm{proc}}$ curve, the overestimated reference makes the normalized runtimes of the larger problems ($512^3$ and $1024^3$) look artificially small—producing the apparent flattening of the slope.

Put differently, the downward trend in the curve does not indicate that the algorithm becomes more efficient at large $N_{\mathrm{proc}}$. Instead, it reflects the growing penalty of load imbalance in the small-scale reference configuration. While this skews the normalized representation, the unnormalized timing data still show the expected, nearly linear increase with problem size. Crucially, in the large production runs (e.g., $1024^3$ at high $N_{\mathrm{proc}}$), the effect of load imbalance is reduced because each process handles a sufficiently large, statistically representative particle sample, resulting in more stable and efficient scaling.

In conclusion, the MPI-parallel implementation delivers strong and scalable performance suitable for large-volume reconstructions. It maintains good strong-scaling behavior up to at least 1024 processes for medium-sized problems ($512^3$) and achieves efficient scaling up to 2048 processes for the large problem ($1024^3$). Although parallel efficiency inevitably drops at very high process counts for a fixed small problem due to communication overheads and load imbalance, performance on large problem sizes—the central focus of this work—remains excellent. The nearly linear dependence of runtime on problem size at fixed parallel resources ensures predictable computational costs for even larger simulations. Collectively, these scaling characteristics satisfy and even surpass the demands of performing full Bayesian reconstructions of initial conditions for next-generation galaxy surveys within realistic computational budgets.

\subsection{Memory Usage}
\label{sec:memoryusage}

In large-scale cosmological reconstructions, available memory typically sets the upper limit on the number of particles that can be used. Moving from a shared-memory (OpenMP) setup to a distributed-memory (MPI) framework is intended precisely to mitigate this limitation by distributing the global state over multiple compute nodes. In this section, we assess the memory efficiency of our implementation.

Figure~\ref{fig:Memory} shows the memory consumption per MPI process for the test setups described in Section~\ref{sec:speed}. The left panel reports strong-scaling memory efficiency: for a fixed global problem size ($256^3$, $512^3$, and $1024^3$), the memory used by each process decreases almost ideally as the number of processes is increased. The tight overlap of the curves indicates that this behavior is essentially independent of the overall problem size.

The right panel plots memory per process against total problem size at fixed numbers of MPI processes. For any given $N_{\mathrm{proc}}$, the memory per process grows linearly with the total number of particles. This is demonstrated by the curves for different $N_{\mathrm{proc}}$ nearly coinciding and following closely the ideal dashed line, which indicates perfect linear scaling extrapolated from the $256^3$ reference. The excellent agreement shows that our domain decomposition strategy incurs negligible overhead and that the memory cost associated with a given subvolume is highly predictable.

A small but systematic feature is that all measured curves fall slightly below the ideal line. This weak departure from perfect linearity is most likely caused by the amortization of fixed or slowly growing overheads—such as allocator metadata, communication buffer bookkeeping, or auxiliary data structures—whose relative impact decreases as the per-process workload grows.

This near-perfect linear and predictable scaling is critical for planning future massive reconstruction runs. It enables robust extrapolation of memory demands to substantially larger problems, ensuring that simulations with up to $4096^3$ particles or more can be scheduled with confidence within the available memory budgets of contemporary high-performance computing systems.

\subsubsection{Practical Implications for ELUCID-DESI Reconstruction}

The nearly linear and highly predictable scaling behavior enables us to extrapolate memory needs to larger problem sizes with substantial confidence. We are currently preparing for ELUCID-DESI runs with approximately $4096^3$ particles. For a halo of mass $M_h$, the corresponding Lagrangian radius is given by $R =(3M_h/4\pi\bar{\rho}_m)^{1/3}$, where $\bar{\rho}_m$ is the mean matter density of the universe. For $M_h \geq 10^{12}M_{\odot}$, this radius is typically $\geq 1Mpc$. The $4096^3$ particles samples this scale with approximately 2.5 cells along each dimension, which is sufficient for the CIC assignment scheme to accurately represent the mass distribution. Our measurements indicate that a reconstruction with $4096^3$ particles distributed over 8192 MPI processors on 128 nodes (64 processors per node) would require on the order of 200 GB of memory per node. This projected usage comfortably fits within the capabilities of current high-performance computing nodes. As a concrete example, the Siyuan cluster offers 512 GB of RAM per node (with 938 nodes and 64 processors per node), so a run of this scale would consume roughly 80\% of the available memory, still leaving adequate margin for the operating system, communication buffers, and I/O overhead.

Consequently, the MPI-based parallelization removes the memory bottleneck that would otherwise make such a large reconstruction infeasible in the original shared-memory (OpenMP) implementation. In combination with the computational scaling behavior discussed in Section~\ref{sec:speed}, this demonstrates that our code can carry out Bayesian reconstructions at the scales required by our upcoming ELUCID-DESI reconstruction, while maintaining predictable, controllable resource demands and thereby ensuring robustness and practicality in real computing environments.

\subsection{Reconstruction Accuracy}
\label{sec:accuracy}

The $\chi^2_{\omega}$ statistic used throughout this section is defined as the weighted sum of squared residuals between the observed and model-predicted density fields (see Equation~\ref{func:chi2}), normalized by the observational uncertainty $\sigma_p(\mathbf{x})$. In practice, $\chi^2_{\omega}$ measures the goodness of fit: smaller values indicate better agreement between the model and the data. When $\chi^2_{\omega}$ drops below $\sim0.01$, the chain is considered to have reached a high-probability region of the posterior, corresponding to a faithful reconstruction of the initial density field.

After examining the computational performance of our MPI-parallelized code, we next focus on its reconstruction accuracy. We assess this accuracy at both low ($256^3$) and high ($1024^3$) resolutions by comparing the reconstructed initial fields to the known true initial conditions from simulations. For the low-resolution tests, we use data from \textsc{Gadget-4} and the \textsc{Quijote} simulation suite \citep{Quijote_sims}, whereas the high-resolution test relies on a simulation from the Kun suite \citep{Chen_2025}.

\begin{figure}
    \centering
    \includegraphics[width=0.9\columnwidth]{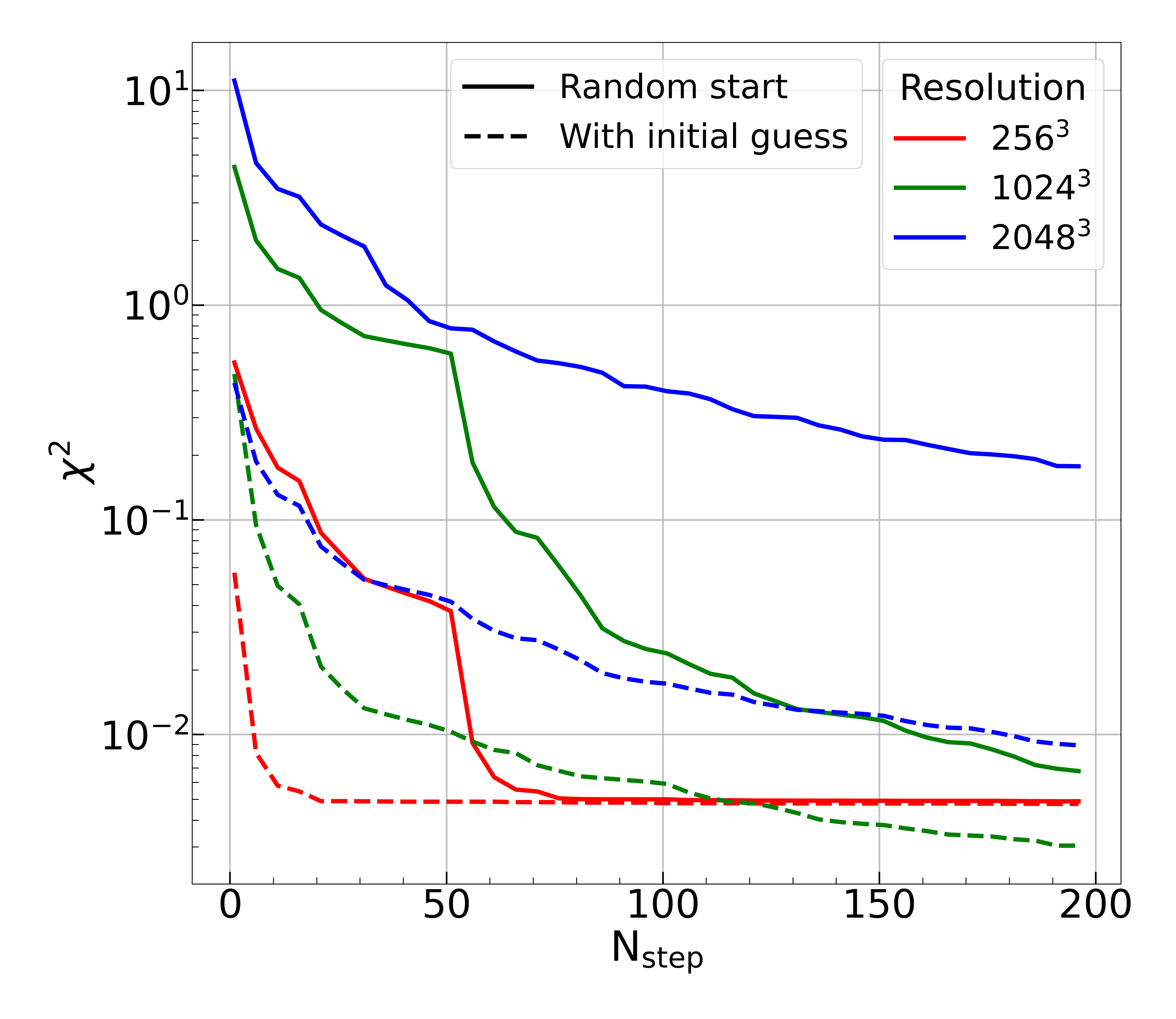}
    \caption{$\chi^2_{\omega}$ values as a function of the number of steps N. Results are shown for chains started from a random field (solid curves) and from the guess module (dashed curves) for three problem sizes ($256^3$, $1024^3$, $2048^3$). In all cases, the guess module delivers a better initial condition (lower starting $\chi^2_{\omega}$) and yields faster convergence. This speed-up becomes increasingly significant as the problem size grows.}
    \label{fig:initial_guess_module}
\end{figure}

\begin{figure*}
	\includegraphics[width=2.0\columnwidth]{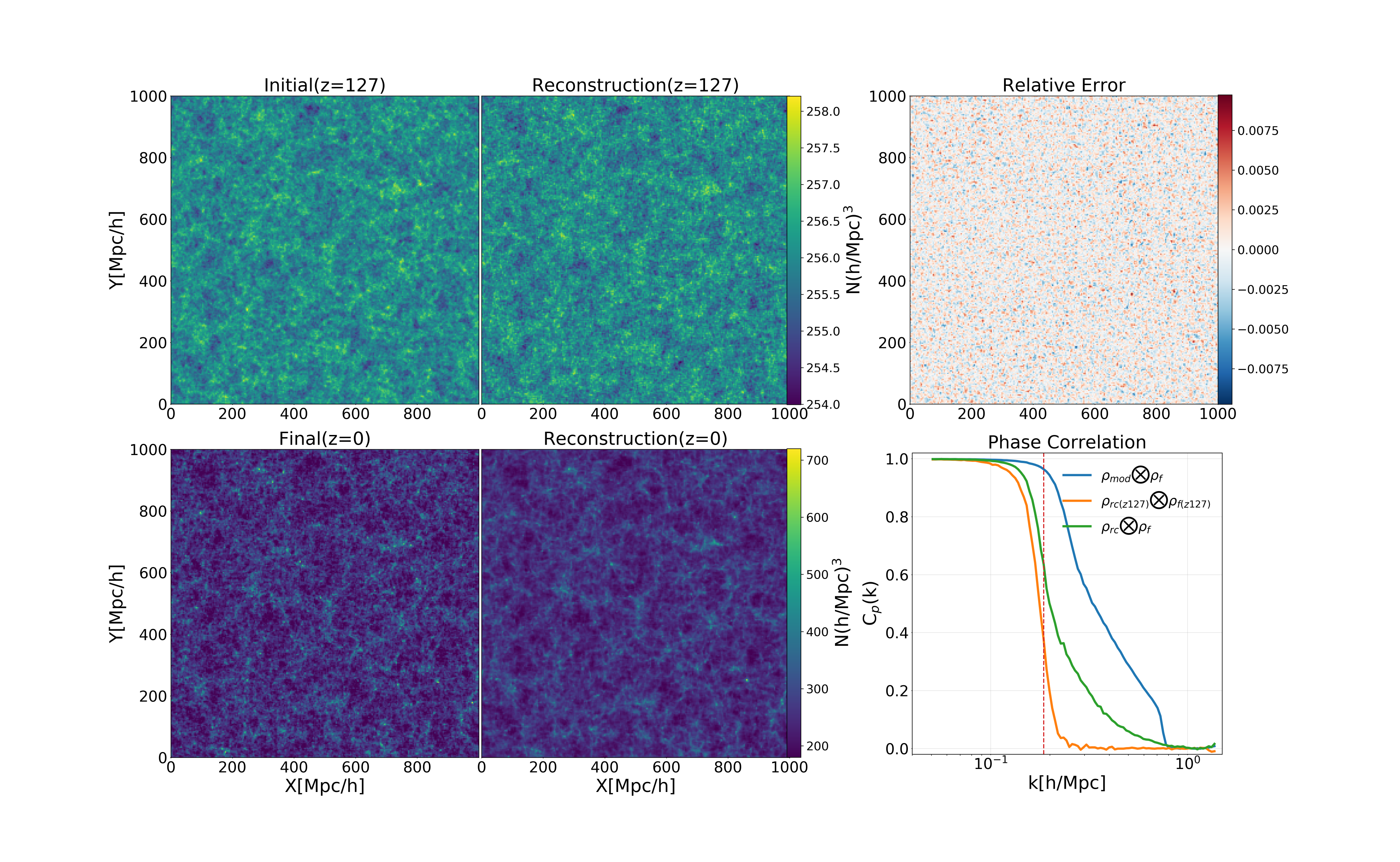}
    \centering
    \caption{Visual and quantitative assessment of a $256^3$ reconstruction. \textbf{Top left:} True initial density field at $z_{\mathrm{ini}}$. \textbf{Top middle:} Reconstructed initial density field. \textbf{Top right:} Relative error map, showing no coherent large-scale bias. \textbf{Bottom left:} Final density field at $z=0$ evolved from the true initial conditions (the reconstruction target). \textbf{Bottom middle:} Final density field at $z=0$ evolved from the reconstructed initial conditions. \textbf{Bottom right:} Phase-correlation coefficient $r(k)$ of Fourier phases. The blue line shows $C_p(k)$ between the smoothed final field ($\rho_{\mathrm{mod}}$) and the true final field. The orange line shows $C_p(k)$ between the reconstructed initial field and the true initial field. The green line shows $C_p(k)$ between the final field evolved from the reconstruction ($\rho_{\mathrm{rc}}$, after CIC assignment) and the true final field. The vertical dashed line marks the theoretical characteristic scale $k_c \approx 1.88/R_s^{0.94}$ from \citet{Wang_2013}.}
    \label{fig:L1000Nd256}
\end{figure*}

\subsubsection{Low-Resolution Tests ($256^3$)}
\label{sec:accuracy_lowres}

We begin by validating the core functionality and convergence of the code at a manageable resolution of $256^3$ particles, using the parameters $N_{\mathrm{pmstep}}=10$, $R_{\mathrm{pm}}=0.75 l_c$, and $R_s=3.0 l_c$, where $l_c \equiv L/Nd $ is the length of the grid cell size. Under these settings, we run the HMCMC sampler for 200 iterations on a $256^3$ density field.

We note that the absolute value of $\chi^2_{\omega}$ at convergence depends on the specific parameter settings (e.g., smoothing scale $R_s$, PM step  $N_{\mathrm{pmstep}}$). The key indicator of convergence is that $\chi^2_{\omega}$ has stabilized and no longer exhibits a systematic decreasing trend. Once this condition is met, the accuracy of the reconstruction has been extensively validated in previous ELUCID work \citep{Wang_2013,Wang_2014,Wang_2016} through comparisons with true initial conditions. In our tests, stable $\chi^2_{\omega}$ values consistently yield unbiased recovery of the density field, power spectrum, and phase correlations (see Figs.~\ref{fig:L1000Nd256} and~\ref{fig:1024pk}).

Convergence is assessed through the evolution of the $\chi^2_{\omega}$ statistic (Equation~\ref{func:chi2}). As indicated by the red solid curve in Figure~\ref{fig:initial_guess_module} (corresponding to a run initialized from a random field), $\chi^2_{\omega}$ rapidly decreases from an initial value of $0.54$ to below $0.01$ within the first 55 iterations, and settles to a stable value of $\sim0.005$ by iteration 80. This trend demonstrates that the HMCMC sampler quickly reaches the high-probability region of the posterior.

A detailed visual and quantitative evaluation of the reconstruction performance is provided in Figure~\ref{fig:L1000Nd256}. The top row shows two-dimensional projections of the 3D density fields, obtained by integrating along one Cartesian axis. The left and middle panels display the true and reconstructed initial density fields, respectively, illustrating that the underlying structure is well recovered. The top-right panel shows the relative error map, which displays no coherent large-scale patterns, indicating that the reconstruction is unbiased. The bottom row shows the corresponding final density fields at $z=0$: the left panel derived from the true initial conditions and the right panel from evolving the reconstructed field. Their close similarity confirms that the forward model accurately reproduces the target distribution.

We further characterize the reconstruction accuracy with the phase correlation of the Fourier phases between the true and reconstructed initial fields, $C_p(k)$, shown in the bottom-right panel. The correlation is unity on large scales and decreases towards smaller scales, as anticipated. The characteristic wavenumber $k_c$ at which $C_p(k)$ experiences a marked decline is well captured by the empirical relation $k_c \approx 1.88 / R_s^{0.94}$ from \citet{Wang_2013}, indicated by the vertical dashed line. This consistency verifies that the smoothing scale $R_s$ effectively controls the scale up to which phase information is recovered, and that our results align with previous foundational work based on OpenMP code.

\begin{figure*}
	\includegraphics[width=1.8\columnwidth]{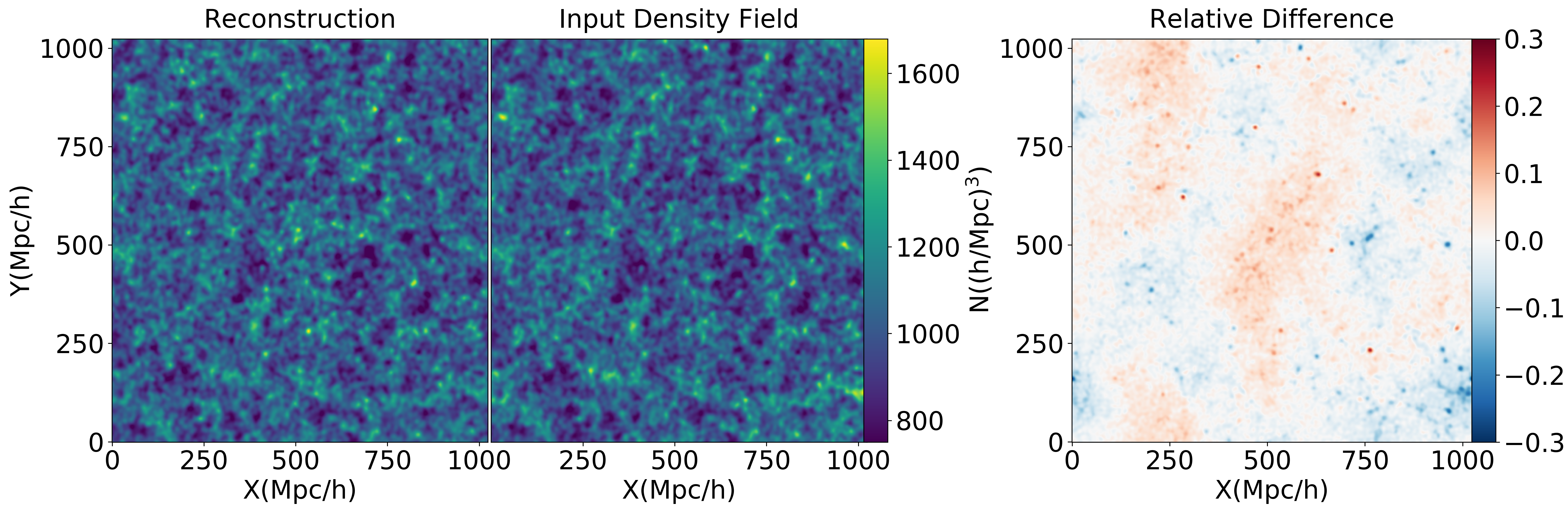}
    \caption{Final density fields at $z=0$ for the $1024^3$ test case. \textbf{Left:} Field obtained by evolving the reconstructed initial conditions with \textsc{Gadget-4}. \textbf{Middle:} Original input field from the Kun simulation. \textbf{Right:} The relative error map. The overall cosmic web structure is faithfully reproduced.}
    \label{fig:L1000Nd1024}
\end{figure*}

\subsubsection{Higher-Resolution Tests ($1024^3$)}
\label{sec:accuracy_highres}

We next evaluate the code at a higher resolution, testing the reconstruction on a density field discretized on a $1024^3$ grid and derived from the Kun simulation. Using parameters $N_{\mathrm{pmstep}}=10$, $R_{\mathrm{pm}}=2.0 l_c$, and $R_s=5.0 l_c$, we perform 200 HMCMC iterations, after which $\chi^2_{\omega}$ converges to a stable value of 0.0066. To provide more fair assessment for the final reconstruction simulation, the reconstructed initial conditions are then forward-evolved to $z=0$ using \textsc{Gadget-4} for direct comparison with the original input field.

A visual inspection of the resulting density fields (Figure~\ref{fig:L1000Nd1024}) shows that the large-scale cosmic web—voids, filaments, and clusters—is well recovered, though some mismatch in the amplitudes of the highest-density clusters persists. In contrast to the relative error map in the upper-right panel of Fig.~\ref{fig:L1000Nd256}, the map shown in the right panel of Fig.~\ref{fig:L1000Nd1024} exhibits a weak but discernible directional dependence, with slightly larger discrepancies along the y-direction. Nevertheless, from our tests comparing the true and reconstructed initial density fields at redshift $z=127$, as well as the final density field at redshift $z=0$, this anisotropy does not appear. We therefore conclude that, despite having used FastPM with 10 time steps to model structure formation, a small mismatch remains when the outcome is compared with the \textsc{Gadget-4} simulation, which is likely the primary source of this directional dependence.

\begin{figure}
\centering
	\includegraphics[width=0.9\columnwidth]{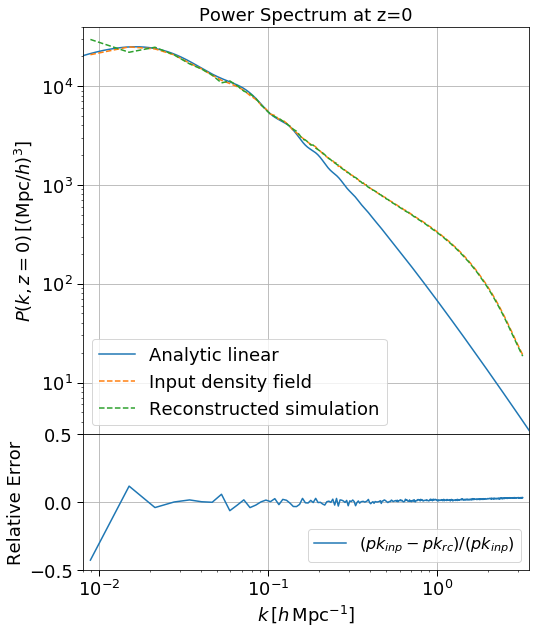}
    \caption{Power spectrum comparison for the $1024^3$ reconstruction. \textbf{Top:} Power spectra of the original (yellow dashed) and reconstructed (green dashed) final density fields, with the linear theory spectrum (black solid) for reference. \textbf{Bottom:} Relative error of the reconstructed power spectrum. Agreement is excellent for $k > 0.05\,h\,\mathrm{Mpc}^{-1}$; the larger errors on the largest scales are consistent with previous HMCMC studies \citep{Wang_2013,Wang_2014}.}
    \label{fig:1024pk}
\end{figure}

To quantify the agreement, we calculate the power spectra of the original and reconstructed final fields (Figure~\ref{fig:1024pk}, top panel). The two power spectra closely match over nearly all wavenumbers, indicating that both the overall amplitude and the scale dependence of the density fluctuations are accurately reconstructed. The lower panel presents the relative difference, which stays below $5\%$ for $k > 0.05\,h\,\mathrm{Mpc}^{-1}$. The larger deviations on the very largest scales (the first few $k$-bins) are a well-known characteristic of HMCMC reconstructions \citep{Wang_2013,Wang_2014}, driven by sample variance from the small number of independent modes in a finite simulation volume.

\begin{figure}
	\includegraphics[width=0.9\columnwidth]{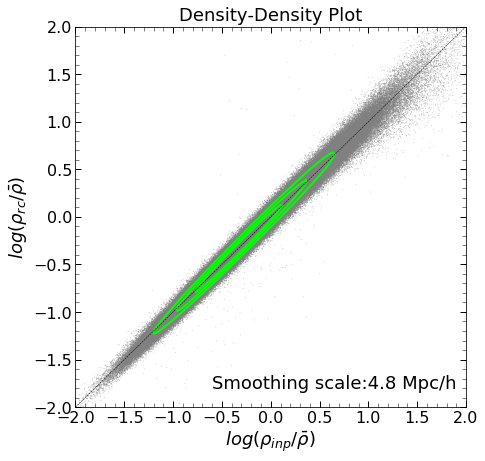}
    \caption{Density-density scatter plot between the original and reconstructed $z=0$ density fields (Gaussian smoothed with $R=4.8\,h^{-1}\mathrm{Mpc}$). Contours enclose $68\%$, $95\%$, and $99\%$ of grid cells. The tight correlation around the one-to-one line (solid black) indicates unbiased reconstruction.}
    \label{fig:Density-Density_Plot}
\end{figure}

Finally, Figure~\ref{fig:Density-Density_Plot} presents a density-density scatter plot between the original and reconstructed fields at $z=0$, smoothed with a $4.8\,h^{-1}\mathrm{Mpc}$ Gaussian kernel. The tight correlation along the one-to-one line (solid black) and the symmetric contours encompassing $67\%$, $95\%$, and $99\%$ of the grid cells indicate unbiased recovery of the density field. The scatter, represented by the width of the contours, quantifies the remaining uncertainty, which is expected given the approximate nature of the FastPM forward model and the finite number of MCMC samples.

\subsubsection{Summary}

The accuracy tests confirm that our MPI-parallelized implementation successfully inherits the reconstruction precision of the original ELUCID algorithm. The code converges robustly, accurately recovers the large-scale structure and power spectrum, and produces statistically unbiased density fields. The observed limitations—smoothing of fine details and increased variance on the largest scales—are understood properties of the method and can be tuned via parameters like $R_s$ and $N_{\mathrm{pmstep}}$. These results validate the code for scientific applications on large-volume datasets.

\subsection{Advantage of the Initial Guess Module}
\label{sec:new_module_result}

The initial guess module introduced in Section~\ref{sec:guess_module} is designed to accelerate HMCMC convergence by providing a high-quality starting point close to the posterior mode. Here, we quantify its effectiveness by comparing reconstruction chains initialized with (i) a random Gaussian field (traditional method) and (ii) the output of our guess module.

Figure~\ref{fig:initial_guess_module} shows the evolution of the $\chi^2_{\omega}$ values as a function of the number of steps N for three problem sizes with $256^3$, $1024^3$, and $2048^3$ particles. In all cases, chains initialized with the guess module (dashed lines) begin with a significantly lower $\chi^2_{\omega}$ than their random-initialized counterparts (solid lines). More importantly, they exhibit a steeper initial descent and reach convergence in markedly fewer iterations.

\begin{table*}
    \centering
        \caption{Computational savings from the initial guess module. \textit{Head-start} indicates how many HMCMC steps (and corresponding core hours) a random-start chain needs to reach the initial $\chi^2_{\omega}$ of the guess-start chain. \textit{Burn-in reduction} is the number of steps (and core hours) saved before convergence. \textit{Baseline Time} lists the estimated total steps (and core hours) required for convergence starting from a random initial field. \textit{Reduction fraction} indicates the proportion of the baseline time saved by the initial guess module. Savings increase significantly with problem size.}
    \begin{tabular}{lcccc}
        \hline
        \textbf{Particles} & \textbf{Head-start (steps)} & \textbf{Burn-in reduction (steps)} & \textbf{Baseline Time (steps)} & \textbf{Reduction fraction} \\ 
        & \textbf{(core hours)} & \textbf{(core hours)} & \textbf{(core hours)} & \textbf{(\%)}\\
        \hline
        $256^3$  & 27 (27.8) & 53 (54.3) & 56 (57.4)& 94.6\\
        $1024^3$ & 53 (3,747) & 106 (7,494) & 160 (11,312) & 66.3\\ 
        $2048^3$ & 90 (72,540) & $\sim$180 (145,080) &  $\sim$360 (290,160) & 50.0\\
        $4096^3$ & $\sim$180 (1,160,640) & $\sim$360 (2,321,280) & $\sim$800 (5,158,400) & 45.0\\
        \hline
    \end{tabular}
    \label{tab:stepandhours}
\end{table*}

We quantify the acceleration in two ways (see Table~\ref{tab:stepandhours}):

\begin{itemize}
    \item \textbf{Head-start advantage} 
\end{itemize}
The initial $\chi^2_{\omega}$ of the guess-initialized chain is equivalent to that achieved by the random-initialized chain only after a substantial number of steps (e.g., 27 steps for $256^3$, 53 steps for $1024^3$ and 90 steps for $2048^3$). This represents a more efficient starting point in the sampling process.

\begin{itemize}
    \item \textbf{Reduction in burn-in} 
\end{itemize}
We define the burn-in period as the number of iterations it takes for the chain’s $\chi^2_{\omega}$ to drop below $1\%$ of its starting value for the first time. Under this definition, the guess-initialized chain converges substantially more rapidly. In the $256^3$ test, it reaches convergence 53 iterations sooner than the randomly initialized chain; in the $1024^3$ test, 106 iterations sooner.

Crucially, the \textit{relative} advantage of the module grows as the problem size increases. This favorable scaling makes it especially useful for the large-volume reconstructions considered in this work. The module introduces only negligible computational overhead (just a single, fast, approximate inversion) while delivering order-of-magnitude reductions in the cost of the subsequent, expensive HMCMC sampling.

These computational gains convert directly into substantial savings in core hours, which become striking at large problem sizes (Table~\ref{tab:stepandhours}). Extrapolating from the $256^3$ and $1024^3$ examples, we estimate that for a $2048^3$ run the guess module could save roughly 145,000 core hours. For our planned ELUCID-DESI simulation with $4096^3$ particles, the savings increase to about 2,300,000 core hours. Based on our scaling tests, we estimate that the $4096^3$ DESI-BGS reconstruction will require approximately $800$ HMCMC steps to reach convergence (starting from the guess module). This corresponds to roughly 5,000,000 core-hours for the baseline chain, and approximately 2,700,000 core-hours after the $45\%$ reduction from the initial guess module.

In conclusion, the initial guess module achieves its intended purpose: it dramatically shortens the burn-in phase of the HMCMC sampler, producing enormous computational savings without degrading the final reconstruction accuracy (which remains the same once the chains have converged). This, in turn, renders rigorous Bayesian reconstruction practical for future survey-scale datasets.

\section{Discussion and Conclusion}
\label{sec:discussion_conclusion}


In this study, aiming at performing the future ELUCID-DESI reconstruction simulations, we have introduced a significant computational advance for the Bayesian reconstruction of the initial condition of the cosmic density field. By re-implementing the HMCMC algorithm in a distributed-memory, MPI-parallel environment and equipping it with an intelligent initial-guess module, we overcome the two main obstacles that previously hindered its application to next-generation survey volumes: memory limitations and excessive computational expense.

The strength of our MPI implementation stems from an effective domain-decomposition strategy. Although we see the expected drop in Processor-number scaling efficiency at very high process counts, performance remains excellent for the large problem sizes that are our primary focus. The code achieves nearly strong scaling (Fig.~\ref{fig:Speed}), indicating that it can make efficient use of additional computational resources to handle larger volumes. This capability is vital for analyzing data from surveys such as CSST and DESI. Furthermore, the almost perfectly linear scaling of memory consumption with problem size (Fig.~\ref{fig:Memory}) permits reliable extrapolation to even larger configurations, turning a $4096^3$ reconstruction into a realistic and schedulable computational task.

The initial-guess module dramatically boosts efficiency. By shortening the HMCMC burn-in phase by tens to hundreds of steps (Fig.~\ref{fig:initial_guess_module}, Table~\ref{tab:stepandhours}), it yields order-of-magnitude savings in total CPU time. Notably, its impact \textit{improves} as the problem size grows, making it highly complementary to our parallelization strategy. The guess field acts as an effective “pre-conditioner” for the sampler, rapidly steering it into the high-likelihood region of the extremely high-dimensional parameter space.

The estimated $\sim$ 2,700,000 core-hours for the $4096^3$ DESI-BGS reconstruction, while substantial, is within the capabilities of modern HPC facilities. For example, the Siyuan cluster used in this work (938 nodes, each with 512 GB RAM) can complete such a production run in a matter of weeks. We therefore consider the ELUCID-DESI reconstruction to be a practical undertaking.


Although we have addressed major scalability issues—making the method suitable for our ELUCID-DESI project—there are still several avenues for further development to support even larger data volumes or more demanding precision in reconstruction.

\textbf{Forward Model Accuracy:} For applications demanding very high accuracy on quasi-linear scales, it would be advantageous to provide a more accurate forward solver as an optional component. The modular design of our framework is intended to make such upgrades straightforward.

\textbf{Optimization for Heterogeneous Architectures:} Additional performance gains may be realized by offloading the most computationally intensive parts—the particle–mesh force calculation and FFTs—to GPUs using a hybrid MPI+OpenACC approach. Moreover, a hybrid MPI/OpenMP strategy could be employed to reduce the number of MPI ranks and mitigate all-to-all communication overhead, particularly for extremely large process counts.

\textbf{Integration with Real Survey Data:} Applying this method to real galaxy survey data will require a careful treatment of the survey mask, redshift-space distortions, and galaxy bias modeling. For ELUCID-DESI, in a separate work we are reconstructing the DESI mass, velocity, and tidal fields using a halo-based group finder framework. Other complementary techniques, including AI-based approaches, have also been developed \citep[e.g.,][]{Wang_2024, 2025ApJS..280...53S}.

\textbf{Enhancing the Guess Module:} The guess module could likely be further improved by embedding a fast nonlinear approximation or a lightweight machine-learning emulator, which could reduce the burn-in period even more. 
Notably, while the absolute computational savings increase dramatically with problem size—from 54 core hours for $256^3$ to over 2,300,000 core hours for $4096^3$—the reduction fraction relative to the baseline time decreases, from 94.6\% at $256^3$ to 45\% at $4096^3$. This trend reflects the increasing complexity of the posterior landscape at larger volumes, where a linear approximation alone becomes less effective at capturing the full nonlinear dynamics. Nevertheless, the absolute savings continue to grow, and even a modest fractional reduction translates into massive gains in absolute terms for the target ELUCID-DESI reconstruction. Future improvements to the guess module, such as incorporating nonlinear corrections, could further push the reduction fraction upward, particularly for the largest scales.

\textbf{Limitations and future validation at the target scale.} The largest reconstruction presented in this work ($2048^3$) is 8 times smaller in volume than the ELUCID-DESI target ($4096^3$). While our scaling tests confirm linear memory scaling and acceptable runtime behavior, we acknowledge that small-scale, high-resolution physics issues may not be fully captured in the current performance-oriented validation. The full $4096^3$ reconstruction on real DESI data is planned as the next phase of the ELUCID-DESI project and will be reported in future work. The present study focuses on the MPI code's performance and scalability; we have therefore confined our validation to scales that are computationally manageable while remaining representative of the expected scaling behavior.

In summary, we have developed and validated a scalable, efficient computational framework that makes rigorous Bayesian reconstruction of the initial density field feasible at the scale of next-generation cosmological surveys. The core of this framework is a high-performance, MPI-parallelized implementation of the HMCMC algorithm that overcomes previous memory barriers. Augmented with a novel initial guess module that dramatically reduces computational cost—with savings increasing with problem size—this tool is now poised to maximize the scientific return from missions like \textit{Euclid}, LSST, CSST and DESI. 

\section*{Acknowledgements}

We thank the referee Adrian Jenkins for his careful reading and constructive suggestions that significantly improved the presentation of this paper. This work is supported by the National Key R\&D Program of China (2023YFA1607800, 2023YFA1607801, 2023YFA1607804),  “the Fundamental Research Funds for the Central Universities”, 111 project No. B20019, the National Natural Science Foundation of China (Grant No. 12273088, 12595312), and Shanghai Natural Science Foundation, grant No.19ZR1466800. We acknowledge the science research grants from the China Manned Space Project with Nos. CMS-CSST-2021-A02 \& CMS-CSST-2025-A04. This project is also supported in part by Office of Science and Technology, Shanghai Municipal Government (grant Nos. 24DX1400100, ZJ2023-ZD-001). HYW is supported by the New Cornerstone Science Foundation through the XPLORER PRIZE. F.S. acknowledges the support from the State Key Laboratory of Dark Matter Physics and the Young Data Scientist Program of the China National Astronomical Data Center (No.NADC2025YDS-01). The computations in this paper were run on the Gravity and Siyuan Supercomputer at Shanghai Jiao Tong University. Thanks to Yu Feng's FastPM code. 

\section*{Data Availability}
 
The data underlying this article will be shared on reasonable request with
the corresponding author.


\bibliographystyle{mnras}
\bibliography{example} 

@ARTICLE{2025ApJS..280...53S,
       author = {{Shi}, Feng and {Wang}, Zitong and {Yang}, Xiaohu and {Gu}, Yizhou and {Wei}, Chengliang and {Li}, Ming and {Han}, Jiaxin and {Ding}, Zhejie and {Wang}, Huiyuan and {Zhang}, Youcai and {Hong}, Wensheng and {Wang}, Yirong and {Li}, Xiao-dong},
        title = "{DarkAI: Reconstructing the Density, Velocity, and Tidal Fields of Dark Matter from a DESI-like Bright Galaxy Sample}",
      journal = {\apjs},
         year = 2025,
        month = oct,
       volume = {280},
       number = {2},
          eid = {53},
        pages = {53},
          doi = {10.3847/1538-4365/adfa26},
archivePrefix = {arXiv},
       eprint = {2501.12621},
 primaryClass = {astro-ph.CO},
       adsurl = {https://ui.adsabs.harvard.edu/abs/2025ApJS..280...53S}
}

@ARTICLE{Hahn_2023,
       author = {{Hahn}, ChangHoon and {Wilson}, Michael J. and {Ruiz-Macias}, Omar and {Cole}, Shaun and {Weinberg}, David H. and {Moustakas}, John and {Kremin}, Anthony and {Tinker}, Jeremy L. and {Smith}, Alex and {Wechsler}, Risa H. and {Ahlen}, Steven and {Alam}, Shadab and {Bailey}, Stephen and {Brooks}, David and {Cooper}, Andrew P. and {Davis}, Tamara M. and {Dawson}, Kyle and {Dey}, Arjun and {Dey}, Biprateep and {Eftekharzadeh}, Sarah and {Eisenstein}, Daniel J. and {Fanning}, Kevin and {Forero-Romero}, Jaime E. and {Frenk}, Carlos S. and {Gazta{\~n}aga}, Enrique and {A Gontcho}, Satya Gontcho and {Guy}, Julien and {Honscheid}, Klaus and {Ishak}, Mustapha and {Juneau}, St{\'e}phanie and {Kehoe}, Robert and {Kisner}, Theodore and {Lan}, Ting-Wen and {Landriau}, Martin and {Le Guillou}, Laurent and {Levi}, Michael E. and {Magneville}, Christophe and {Martini}, Paul and {Meisner}, Aaron and {Myers}, Adam D. and {Nie}, Jundan and {Norberg}, Peder and {Palanque-Delabrouille}, Nathalie and {Percival}, Will J. and {Poppett}, Claire and {Prada}, Francisco and {Raichoor}, Anand and {Ross}, Ashley J. and {Gaines}, Sasha and {Saulder}, Christoph and {Schlafly}, Eddie and {Schlegel}, David and {Sierra-Porta}, David and {Tarle}, Gregory and {Weaver}, Benjamin A. and {Y{\`e}che}, Christophe and {Zarrouk}, Pauline and {Zhou}, Rongpu and {Zhou}, Zhimin and {Zou}, Hu},
        title = "{The DESI Bright Galaxy Survey: Final Target Selection, Design, and Validation}",
      journal = {\aj},
         year = 2023,
        month = jun,
       volume = {165},
       number = {6},
          eid = {253},
        pages = {253},
          doi = {10.3847/1538-3881/accff8},
archivePrefix = {arXiv},
       eprint = {2208.08512},
 primaryClass = {astro-ph.CO},
       adsurl = {https://ui.adsabs.harvard.edu/abs/2023AJ....165..253H}
}

@ARTICLE{Myers_2023,
       author = {{Myers}, Adam D. and {Moustakas}, John and {Bailey}, Stephen and {Weaver}, Benjamin A. and {Cooper}, Andrew P. and {Forero-Romero}, Jaime E. and {Abolfathi}, Bela and {Alexander}, David M. and {Brooks}, David and {Chaussidon}, Edmond and {Chuang}, Chia-Hsun and {Dawson}, Kyle and {Dey}, Arjun and {Dey}, Biprateep and {Dhungana}, Govinda and {Doel}, Peter and {Fanning}, Kevin and {Gazta{\~n}aga}, Enrique and {Gontcho A Gontcho}, Satya and {Gonzalez-Morales}, Alma X. and {Hahn}, ChangHoon and {Herrera-Alcantar}, Hiram K. and {Honscheid}, Klaus and {Ishak}, Mustapha and {Karim}, Tanveer and {Kirkby}, David and {Kisner}, Theodore and {Koposov}, Sergey E. and {Kremin}, Anthony and {Lan}, Ting-Wen and {Landriau}, Martin and {Lang}, Dustin and {Levi}, Michael E. and {Magneville}, Christophe and {Napolitano}, Lucas and {Martini}, Paul and {Meisner}, Aaron and {Newman}, Jeffrey A. and {Palanque-Delabrouille}, Nathalie and {Percival}, Will and {Poppett}, Claire and {Prada}, Francisco and {Raichoor}, Anand and {Ross}, Ashley J. and {Schlafly}, Edward F. and {Schlegel}, David and {Schubnell}, Michael and {Tan}, Ting and {Tarle}, Gregory and {Wilson}, Michael J. and {Y{\`e}che}, Christophe and {Zhou}, Rongpu and {Zhou}, Zhimin and {Zou}, Hu},
        title = "{The Target-selection Pipeline for the Dark Energy Spectroscopic Instrument}",
      journal = {\aj},
         year = 2023,
        month = feb,
       volume = {165},
       number = {2},
          eid = {50},
        pages = {50},
          doi = {10.3847/1538-3881/aca5f9},
archivePrefix = {arXiv},
       eprint = {2208.08518},
 primaryClass = {astro-ph.IM},
       adsurl = {https://ui.adsabs.harvard.edu/abs/2023AJ....165...50M}
}

@ARTICLE{Kauffmann_1993,
       author = {{Kauffmann}, G. and {White}, S.~D.~M. and {Guiderdoni}, B.},
        title = "{The formation and evolution of galaxies within merging dark matter haloes.}",
      journal = {\mnras},
         year = 1993,
        month = sep,
       volume = {264},
        pages = {201-218},
          doi = {10.1093/mnras/264.1.201},
       adsurl = {https://ui.adsabs.harvard.edu/abs/1993MNRAS.264..201K}
}

@ARTICLE{Guo_2011,
       author = {{Guo}, Qi and {White}, Simon and {Boylan-Kolchin}, Michael and {De Lucia}, Gabriella and {Kauffmann}, Guinevere and {Lemson}, Gerard and {Li}, Cheng and {Springel}, Volker and {Weinmann}, Simone},
        title = "{From dwarf spheroidals to cD galaxies: simulating the galaxy population in a {\ensuremath{\Lambda}}CDM cosmology}",
      journal = {\mnras},
         year = 2011,
        month = may,
       volume = {413},
       number = {1},
        pages = {101-131},
          doi = {10.1111/j.1365-2966.2010.18114.x},
archivePrefix = {arXiv},
       eprint = {1006.0106},
 primaryClass = {astro-ph.CO},
       adsurl = {https://ui.adsabs.harvard.edu/abs/2011MNRAS.413..101G}
}

@ARTICLE{Blanton_2005,
       author = {{Blanton}, Michael R. and {Schlegel}, David J. and {Strauss}, Michael A. and {Brinkmann}, J. and {Finkbeiner}, Douglas and {Fukugita}, Masataka and {Gunn}, James E. and {Hogg}, David W. and {Ivezi{\'c}}, {\v{Z}}eljko and {Knapp}, G.~R. and {Lupton}, Robert H. and {Munn}, Jeffrey A. and {Schneider}, Donald P. and {Tegmark}, Max and {Zehavi}, Idit},
        title = "{New York University Value-Added Galaxy Catalog: A Galaxy Catalog Based on New Public Surveys}",
      journal = {\aj},
         year = 2005,
        month = jun,
       volume = {129},
       number = {6},
        pages = {2562-2578},
          doi = {10.1086/429803},
archivePrefix = {arXiv},
       eprint = {astro-ph/0410166},
 primaryClass = {astro-ph},
       adsurl = {https://ui.adsabs.harvard.edu/abs/2005AJ....129.2562B}
}

@ARTICLE{Kitaura_2013,
       author = {{Kitaura}, F.-S.},
        title = "{The initial conditions of the universe from constrained simulations.}",
      journal = {\mnras},
         year = 2013,
        month = feb,
       volume = {429},
        pages = {L84-L88},
          doi = {10.1093/mnrasl/sls029},
archivePrefix = {arXiv},
       eprint = {1203.4184},
 primaryClass = {astro-ph.CO},
       adsurl = {https://ui.adsabs.harvard.edu/abs/2013MNRAS.429L..84K}
}

@ARTICLE{WangYR_2024,
       author = {{Wang}, Yirong and {Yang}, Xiaohu and {Gu}, Yizhou and {Xu}, Xiaoju and {Xu}, Haojie and {Wang}, Yuyu and {Katsianis}, Antonios and {Han}, Jiaxin and {He}, Min and {Zheng}, Yunliang and {Li}, Qingyang and {Wang}, Yaru and {Hong}, Wensheng and {Wang}, Jiaqi and {Tan}, Zhenlin and {Zou}, Hu and {Lange}, Johannes Ulf and {Hahn}, ChangHoon and {Behroozi}, Peter and {Aguilar}, Jessica Nicole and {Ahlen}, Steven and {Brooks}, David and {Claybaugh}, Todd and {Cole}, Shaun and {de la Macorra}, Axel and {Dey}, Biprateep and {Doel}, Peter and {Forero-Romero}, Jaime E. and {Honscheid}, Klaus and {Kehoe}, Robert and {Kisner}, Theodore and {Lambert}, Andrew and {Manera}, Marc and {Meisner}, Aaron and {Miquel}, Ramon and {Moustakas}, John and {Nie}, Jundan and {Poppett}, Claire and {Rezaie}, Mehdi and {Rossi}, Graziano and {Sanchez}, Eusebio and {Schubnell}, Michael and {Tarl{\'e}}, Gregory and {Weaver}, Benjamin Alan and {Zhou}, Zhimin},
        title = "{Measuring the Conditional Luminosity and Stellar Mass Functions of Galaxies by Combining the Dark Energy Spectroscopic Instrument Legacy Imaging Surveys Data Release 9, Survey Validation 3, and Year 1 Data}",
      journal = {\apj},
         year = 2024,
        month = aug,
       volume = {971},
       number = {1},
          eid = {119},
        pages = {119},
          doi = {10.3847/1538-4357/ad5294},
archivePrefix = {arXiv},
       eprint = {2312.17459},
 primaryClass = {astro-ph.GA},
       adsurl = {https://ui.adsabs.harvard.edu/abs/2024ApJ...971..119W}
}

@ARTICLE{Yang_2012,
       author = {{Yang}, Xiaohu and {Mo}, H.~J. and {van den Bosch}, Frank C. and {Zhang}, Youcai and {Han}, Jiaxin},
        title = "{Evolution of the Galaxy-Dark Matter Connection and the Assembly of Galaxies in Dark Matter Halos}",
      journal = {\apj},
         year = 2012,
        month = jun,
       volume = {752},
       number = {1},
          eid = {41},
        pages = {41},
          doi = {10.1088/0004-637X/752/1/41},
archivePrefix = {arXiv},
       eprint = {1110.1420},
 primaryClass = {astro-ph.CO},
       adsurl = {https://ui.adsabs.harvard.edu/abs/2012ApJ...752...41Y}
}

@ARTICLE{Yang_2008,
       author = {{Yang}, Xiaohu and {Mo}, H.~J. and {van den Bosch}, Frank C.},
        title = "{Galaxy Groups in the SDSS DR4. II. Halo Occupation Statistics}",
      journal = {\apj},
         year = 2008,
        month = mar,
       volume = {676},
       number = {1},
        pages = {248-261},
          doi = {10.1086/528954},
archivePrefix = {arXiv},
       eprint = {0710.5096},
 primaryClass = {astro-ph},
       adsurl = {https://ui.adsabs.harvard.edu/abs/2008ApJ...676..248Y}
}

@ARTICLE{Yang_2009,
       author = {{Yang}, Xiaohu and {Mo}, H.~J. and {van den Bosch}, Frank C.},
        title = "{Galaxy Groups in the SDSS DR4. III. The Luminosity and Stellar Mass Functions}",
      journal = {\apj},
         year = 2009,
        month = apr,
       volume = {695},
       number = {2},
        pages = {900-916},
          doi = {10.1088/0004-637X/695/2/900},
archivePrefix = {arXiv},
       eprint = {0808.0539},
 primaryClass = {astro-ph},
       adsurl = {https://ui.adsabs.harvard.edu/abs/2009ApJ...695..900Y}
}

@ARTICLE{Zehavi_2005,
       author = {{Zehavi}, Idit and {Zheng}, Zheng and {Weinberg}, David H. and {Frieman}, Joshua A. and {Berlind}, Andreas A. and {Blanton}, Michael R. and {Scoccimarro}, Roman and {Sheth}, Ravi K. and {Strauss}, Michael A. and {Kayo}, Issha and {Suto}, Yasushi and {Fukugita}, Masataka and {Nakamura}, Osamu and {Bahcall}, Neta A. and {Brinkmann}, Jon and {Gunn}, James E. and {Hennessy}, Greg S. and {Ivezi{\'c}}, {\v{Z}}eljko and {Knapp}, Gillian R. and {Loveday}, Jon and {Meiksin}, Avery and {Schlegel}, David J. and {Schneider}, Donald P. and {Szapudi}, Istvan and {Tegmark}, Max and {Vogeley}, Michael S. and {York}, Donald G. and {SDSS Collaboration}},
        title = "{The Luminosity and Color Dependence of the Galaxy Correlation Function}",
      journal = {\apj},
         year = 2005,
        month = sep,
       volume = {630},
       number = {1},
        pages = {1-27},
          doi = {10.1086/431891},
archivePrefix = {arXiv},
       eprint = {astro-ph/0408569},
 primaryClass = {astro-ph},
       adsurl = {https://ui.adsabs.harvard.edu/abs/2005ApJ...630....1Z}
}

@ARTICLE{Zehavi_2011,
       author = {{Zehavi}, Idit and {Zheng}, Zheng and {Weinberg}, David H. and {Blanton}, Michael R. and {Bahcall}, Neta A. and {Berlind}, Andreas A. and {Brinkmann}, Jon and {Frieman}, Joshua A. and {Gunn}, James E. and {Lupton}, Robert H. and {Nichol}, Robert C. and {Percival}, Will J. and {Schneider}, Donald P. and {Skibba}, Ramin A. and {Strauss}, Michael A. and {Tegmark}, Max and {York}, Donald G.},
        title = "{Galaxy Clustering in the Completed SDSS Redshift Survey: The Dependence on Color and Luminosity}",
      journal = {\apj},
         year = 2011,
        month = jul,
       volume = {736},
       number = {1},
          eid = {59},
        pages = {59},
          doi = {10.1088/0004-637X/736/1/59},
archivePrefix = {arXiv},
       eprint = {1005.2413},
 primaryClass = {astro-ph.CO},
       adsurl = {https://ui.adsabs.harvard.edu/abs/2011ApJ...736...59Z}
}

@ARTICLE{Zheng_2005,
       author = {{Zheng}, Zheng and {Berlind}, Andreas A. and {Weinberg}, David H. and {Benson}, Andrew J. and {Baugh}, Carlton M. and {Cole}, Shaun and {Dav{\'e}}, Romeel and {Frenk}, Carlos S. and {Katz}, Neal and {Lacey}, Cedric G.},
        title = "{Theoretical Models of the Halo Occupation Distribution: Separating Central and Satellite Galaxies}",
      journal = {\apj},
         year = 2005,
        month = nov,
       volume = {633},
       number = {2},
        pages = {791-809},
          doi = {10.1086/466510},
archivePrefix = {arXiv},
       eprint = {astro-ph/0408564},
 primaryClass = {astro-ph},
       adsurl = {https://ui.adsabs.harvard.edu/abs/2005ApJ...633..791Z}
}

@ARTICLE{Bertschinger_1987,
       author = {{Bertschinger}, Edmund},
        title = "{Path Integral Methods for Primordial Density Perturbations: Sampling of Constrained Gaussian Random Fields}",
      journal = {\apjl},
         year = 1987,
        month = dec,
       volume = {323},
        pages = {L103},
          doi = {10.1086/185066},
       adsurl = {https://ui.adsabs.harvard.edu/abs/1987ApJ...323L.103B}
}

@ARTICLE{Doumler_2013,
       author = {{Doumler}, Timur and {Hoffman}, Yehuda and {Courtois}, H{\'e}l{\`e}ne and {Gottl{\"o}ber}, Stefan},
        title = "{Reconstructing cosmological initial conditions from galaxy peculiar velocities - I. Reverse Zeldovich Approximation}",
      journal = {\mnras},
         year = 2013,
        month = apr,
       volume = {430},
       number = {2},
        pages = {888-901},
          doi = {10.1093/mnras/sts613},
archivePrefix = {arXiv},
       eprint = {1212.2806},
 primaryClass = {astro-ph.CO},
       adsurl = {https://ui.adsabs.harvard.edu/abs/2013MNRAS.430..888D}
}

@ARTICLE{Hoffman_1991,
       author = {{Hoffman}, Yehuda and {Ribak}, Erez},
        title = "{Constrained Realizations of Gaussian Fields: A Simple Algorithm}",
      journal = {\apjl},
         year = 1991,
        month = oct,
       volume = {380},
        pages = {L5},
          doi = {10.1086/186160},
       adsurl = {https://ui.adsabs.harvard.edu/abs/1991ApJ...380L...5H}
}

@ARTICLE{Jasche_2013,
       author = {{Jasche}, Jens and {Wandelt}, Benjamin D.},
        title = "{Bayesian physical reconstruction of initial conditions from large-scale structure surveys}",
      journal = {\mnras},
         year = 2013,
        month = jun,
       volume = {432},
       number = {2},
        pages = {894-913},
          doi = {10.1093/mnras/stt449},
archivePrefix = {arXiv},
       eprint = {1203.3639},
 primaryClass = {astro-ph.CO},
       adsurl = {https://ui.adsabs.harvard.edu/abs/2013MNRAS.432..894J}
}

@ARTICLE{Jing_1998,
       author = {{Jing}, Y.~P. and {Mo}, H.~J. and {B{\"o}rner}, G.},
        title = "{Spatial Correlation Function and Pairwise Velocity Dispersion of Galaxies: Cold Dark Matter Models versus the Las Campanas Survey}",
      journal = {\apj},
         year = 1998,
        month = feb,
       volume = {494},
       number = {1},
        pages = {1-12},
          doi = {10.1086/305209},
archivePrefix = {arXiv},
       eprint = {astro-ph/9707106},
 primaryClass = {astro-ph},
       adsurl = {https://ui.adsabs.harvard.edu/abs/1998ApJ...494....1J}
}

@ARTICLE{Kitaura_2008,
       author = {{Kitaura}, F.~S. and {En{\ss}lin}, T.~A.},
        title = "{Bayesian reconstruction of the cosmological large-scale structure: methodology, inverse algorithms and numerical optimization}",
      journal = {\mnras},
         year = 2008,
        month = sep,
       volume = {389},
       number = {2},
        pages = {497-544},
          doi = {10.1111/j.1365-2966.2008.13341.x},
archivePrefix = {arXiv},
       eprint = {0705.0429},
 primaryClass = {astro-ph},
       adsurl = {https://ui.adsabs.harvard.edu/abs/2008MNRAS.389..497K}
}

@ARTICLE{Klypin_2003,
       author = {{Klypin}, Anatoly and {Hoffman}, Yehuda and {Kravtsov}, Andrey V. and {Gottl{\"o}ber}, Stefan},
        title = "{Constrained Simulations of the Real Universe: The Local Supercluster}",
      journal = {\apj},
         year = 2003,
        month = oct,
       volume = {596},
       number = {1},
        pages = {19-33},
          doi = {10.1086/377574},
archivePrefix = {arXiv},
       eprint = {astro-ph/0107104},
 primaryClass = {astro-ph},
       adsurl = {https://ui.adsabs.harvard.edu/abs/2003ApJ...596...19K}
}

@ARTICLE{Kravtsov_2002,
       author = {{Kravtsov}, Andrey V. and {Klypin}, Anatoly and {Hoffman}, Yehuda},
        title = "{Constrained Simulations of the Real Universe. II. Observational Signatures of Intergalactic Gas in the Local Supercluster Region}",
      journal = {\apj},
         year = 2002,
        month = jun,
       volume = {571},
       number = {2},
        pages = {563-575},
          doi = {10.1086/340046},
archivePrefix = {arXiv},
       eprint = {astro-ph/0109077},
 primaryClass = {astro-ph},
       adsurl = {https://ui.adsabs.harvard.edu/abs/2002ApJ...571..563K}
}

@ARTICLE{Seljak_2017,
       author = {{Seljak}, Uro{\v{s}} and {Aslanyan}, Grigor and {Feng}, Yu and {Modi}, Chirag},
        title = "{Towards optimal extraction of cosmological information from nonlinear data}",
      journal = {\jcap},
         year = 2017,
        month = dec,
       volume = {2017},
       number = {12},
          eid = {009},
        pages = {009},
          doi = {10.1088/1475-7516/2017/12/009},
archivePrefix = {arXiv},
       eprint = {1706.06645},
 primaryClass = {astro-ph.CO},
       adsurl = {https://ui.adsabs.harvard.edu/abs/2017JCAP...12..009S}
}

@ARTICLE{Sousa_2007,
       author = {{Sousa}, S.~G. and {Santos}, N.~C. and {Israelian}, G. and {Mayor}, M. and {Monteiro}, M.~J.~P.~F.~G.},
        title = "{A new code for automatic determination of equivalent widths: Automatic Routine for line Equivalent widths in stellar Spectra (ARES)}",
      journal = {\aap},
         year = 2007,
        month = jul,
       volume = {469},
       number = {2},
        pages = {783-791},
          doi = {10.1051/0004-6361:20077288},
archivePrefix = {arXiv},
       eprint = {astro-ph/0703696},
 primaryClass = {astro-ph},
       adsurl = {https://ui.adsabs.harvard.edu/abs/2007A&A...469..783S}
}

@ARTICLE{van_de_Weygaert_1996,
       author = {{van de Weygaert}, Rien and {Bertschinger}, Edmund},
        title = "{Peak and gravity constraints in Gaussian primordial density fields: An application of the Hoffman-Ribak method}",
      journal = {\mnras},
         year = 1996,
        month = jul,
       volume = {281},
        pages = {84},
          doi = {10.1093/mnras/281.1.84},
archivePrefix = {arXiv},
       eprint = {astro-ph/9507024},
 primaryClass = {astro-ph},
       adsurl = {https://ui.adsabs.harvard.edu/abs/1996MNRAS.281...84V}
}

@ARTICLE{JUST_2024,
       author = {{JUST Team} and {Liu}, Chengze and {Zu}, Ying and {Feng}, Fabo and {Li}, Zhaoyu and {Yu}, Yu and {Bai}, Hua and {Cui}, Xiangqun and {Gu}, Bozhong and {Gu}, Yizhou and {Han}, Jiaxin and {Hou}, Yonghui and {Hu}, Zhongwen and {Ji}, Hangxin and {Jing}, Yipeng and {Li}, Wei and {Qi}, Zhaoxiang and {Tan}, Xianyu and {Tian}, Cairang and {Yang}, Dehua and {Yuan}, Xiangyan and {Zhai}, Chao and {Zhang}, Congcong and {Zhang}, Jun and {Zhang}, Haotong and {Zhang}, Pengjie and {Zhang}, Yong and {Zhao}, Yi and {Zheng}, Xianzhong and {Zhu}, Qingfeng and {Yang}, Xiaohu},
        title = "{The Jiao Tong University Spectroscopic Telescope (JUST) Project}",
      journal = {Astronomical Techniques and Instruments},
         year = 2024,
        month = jan,
       volume = {1},
       number = {1},
        pages = {16-30},
          doi = {10.61977/ati2024008},
archivePrefix = {arXiv},
       eprint = {2402.14312},
 primaryClass = {astro-ph.IM},
       adsurl = {https://ui.adsabs.harvard.edu/abs/2024AstTI...1...16J}
}

@BOOK{Mo_2010,
       author = {{Mo}, Houjun and {van den Bosch}, Frank C. and {White}, Simon},
        title = "{Galaxy Formation and Evolution}",
         year = 2010,
          doi = {10.1017/CBO9780511807244},
       adsurl = {https://ui.adsabs.harvard.edu/abs/2010gfe..book.....M},
       publisher={Cambridge University Press}
}

@ARTICLE{Takada_2014,
       author = {{Takada}, Masahiro and {Ellis}, Richard S. and {Chiba}, Masashi and {Greene}, Jenny E. and {Aihara}, Hiroaki and {Arimoto}, Nobuo and {Bundy}, Kevin and {Cohen}, Judith and {Dor{\'e}}, Olivier and {Graves}, Genevieve and {Gunn}, James E. and {Heckman}, Timothy and {Hirata}, Christopher M. and {Ho}, Paul and {Kneib}, Jean-Paul and {Le F{\`e}vre}, Olivier and {Lin}, Lihwai and {More}, Surhud and {Murayama}, Hitoshi and {Nagao}, Tohru and {Ouchi}, Masami and {Seiffert}, Michael and {Silverman}, John D. and {Sodr{\'e}}, Laerte and {Spergel}, David N. and {Strauss}, Michael A. and {Sugai}, Hajime and {Suto}, Yasushi and {Takami}, Hideki and {Wyse}, Rosemary},
        title = "{Extragalactic science, cosmology, and Galactic archaeology with the Subaru Prime Focus Spectrograph}",
      journal = {\pasj},
         year = 2014,
        month = feb,
       volume = {66},
       number = {1},
          eid = {R1},
        pages = {R1},
          doi = {10.1093/pasj/pst019},
archivePrefix = {arXiv},
       eprint = {1206.0737},
 primaryClass = {astro-ph.CO},
       adsurl = {https://ui.adsabs.harvard.edu/abs/2014PASJ...66R...1T}
}

@ARTICLE{Zhang_2025,
       author = {{Zhang}, Youcai and {Yang}, Xiaohu and {Guo}, Hong and {Wang}, Peng and {Shi}, Feng},
        title = "{Galaxy and halo properties around cosmic filaments from Sloan Digital Sky Survey Data Release 7 and the ELUCID simulation}",
      journal = {\mnras},
         year = 2025,
        month = may,
       volume = {539},
       number = {2},
        pages = {1692-1705},
          doi = {10.1093/mnras/staf611},
archivePrefix = {arXiv},
       eprint = {2504.07367},
 primaryClass = {astro-ph.CO},
       adsurl = {https://ui.adsabs.harvard.edu/abs/2025MNRAS.539.1692Z}
}

@ARTICLE{Wang_2024,
       author = {{Wang}, Zitong and {Shi}, Feng and {Yang}, Xiaohu and {Li}, Qingyang and {Liu}, Yanming and {Li}, Xiaoping},
        title = "{(DarkAI) Mapping the large-scale density field of dark matter using artificial intelligence}",
      journal = {Science China Physics, Mechanics, and Astronomy},
         year = 2024,
        month = jan,
       volume = {67},
       number = {1},
          eid = {219513},
        pages = {219513},
          doi = {10.1007/s11433-023-2192-9},
archivePrefix = {arXiv},
       eprint = {2305.11431},
 primaryClass = {astro-ph.CO},
       adsurl = {https://ui.adsabs.harvard.edu/abs/2024SCPMA..6719513W}
}

@ARTICLE{Zhang_2022,
       author = {{Zhang}, Youcai and {Yang}, Xiaohu and {Guo}, Hong},
        title = "{Galaxy-halo size relation from Sloan Digital Sky Survey Data Release 7 and the ELUCID simulation}",
      journal = {\mnras},
         year = 2022,
        month = dec,
       volume = {517},
       number = {3},
        pages = {3579-3587},
          doi = {10.1093/mnras/stac2934},
archivePrefix = {arXiv},
       eprint = {2210.05215},
 primaryClass = {astro-ph.GA},
       adsurl = {https://ui.adsabs.harvard.edu/abs/2022MNRAS.517.3579Z}
}

@ARTICLE{Zhang_2021,
       author = {{Zhang}, Youcai and {Yang}, Xiaohu and {Guo}, Hong},
        title = "{Galaxy-group (halo) alignments from SDSS DR7 and the ELUCID simulation}",
      journal = {\mnras},
         year = 2021,
        month = jan,
       volume = {500},
       number = {2},
        pages = {1895-1904},
          doi = {10.1093/mnras/staa2356},
archivePrefix = {arXiv},
       eprint = {2008.01381},
 primaryClass = {astro-ph.CO},
       adsurl = {https://ui.adsabs.harvard.edu/abs/2021MNRAS.500.1895Z}
}

@ARTICLE{SDSS_2000,
       author = {{York}, Donald G. and {Adelman}, J. and {Anderson}, Jr., John E. and {Anderson}, Scott F. and {Annis}, James and {Bahcall}, Neta A. and {Bakken}, J.~A. and {Barkhouser}, Robert and {Bastian}, Steven and {Berman}, Eileen and {Boroski}, William N. and {Bracker}, Steve and {Briegel}, Charlie and {Briggs}, John W. and {Brinkmann}, J. and {Brunner}, Robert and {Burles}, Scott and {Carey}, Larry and {Carr}, Michael A. and {Castander}, Francisco J. and {Chen}, Bing and {Colestock}, Patrick L. and {Connolly}, A.~J. and {Crocker}, J.~H. and {Csabai}, Istv{\'a}n and {Czarapata}, Paul C. and {Davis}, John Eric and {Doi}, Mamoru and {Dombeck}, Tom and {Eisenstein}, Daniel and {Ellman}, Nancy and {Elms}, Brian R. and {Evans}, Michael L. and {Fan}, Xiaohui and {Federwitz}, Glenn R. and {Fiscelli}, Larry and {Friedman}, Scott and {Frieman}, Joshua A. and {Fukugita}, Masataka and {Gillespie}, Bruce and {Gunn}, James E. and {Gurbani}, Vijay K. and {de Haas}, Ernst and {Haldeman}, Merle and {Harris}, Frederick H. and {Hayes}, J. and {Heckman}, Timothy M. and {Hennessy}, G.~S. and {Hindsley}, Robert B. and {Holm}, Scott and {Holmgren}, Donald J. and {Huang}, Chi-hao and {Hull}, Charles and {Husby}, Don and {Ichikawa}, Shin-Ichi and {Ichikawa}, Takashi and {Ivezi{\'c}}, {\v{Z}}eljko and {Kent}, Stephen and {Kim}, Rita S.~J. and {Kinney}, E. and {Klaene}, Mark and {Kleinman}, A.~N. and {Kleinman}, S. and {Knapp}, G.~R. and {Korienek}, John and {Kron}, Richard G. and {Kunszt}, Peter Z. and {Lamb}, D.~Q. and {Lee}, B. and {Leger}, R. French and {Limmongkol}, Siriluk and {Lindenmeyer}, Carl and {Long}, Daniel C. and {Loomis}, Craig and {Loveday}, Jon and {Lucinio}, Rich and {Lupton}, Robert H. and {MacKinnon}, Bryan and {Mannery}, Edward J. and {Mantsch}, P.~M. and {Margon}, Bruce and {McGehee}, Peregrine and {McKay}, Timothy A. and {Meiksin}, Avery and {Merelli}, Aronne and {Monet}, David G. and {Munn}, Jeffrey A. and {Narayanan}, Vijay K. and {Nash}, Thomas and {Neilsen}, Eric and {Neswold}, Rich and {Newberg}, Heidi Jo and {Nichol}, R.~C. and {Nicinski}, Tom and {Nonino}, Mario and {Okada}, Norio and {Okamura}, Sadanori and {Ostriker}, Jeremiah P. and {Owen}, Russell and {Pauls}, A. George and {Peoples}, John and {Peterson}, R.~L. and {Petravick}, Donald and {Pier}, Jeffrey R. and {Pope}, Adrian and {Pordes}, Ruth and {Prosapio}, Angela and {Rechenmacher}, Ron and {Quinn}, Thomas R. and {Richards}, Gordon T. and {Richmond}, Michael W. and {Rivetta}, Claudio H. and {Rockosi}, Constance M. and {Ruthmansdorfer}, Kurt and {Sandford}, Dale and {Schlegel}, David J. and {Schneider}, Donald P. and {Sekiguchi}, Maki and {Sergey}, Gary and {Shimasaku}, Kazuhiro and {Siegmund}, Walter A. and {Smee}, Stephen and {Smith}, J. Allyn and {Snedden}, S. and {Stone}, R. and {Stoughton}, Chris and {Strauss}, Michael A. and {Stubbs}, Christopher and {SubbaRao}, Mark and {Szalay}, Alexander S. and {Szapudi}, Istvan and {Szokoly}, Gyula P. and {Thakar}, Anirudda R. and {Tremonti}, Christy and {Tucker}, Douglas L. and {Uomoto}, Alan and {Vanden Berk}, Dan and {Vogeley}, Michael S. and {Waddell}, Patrick and {Wang}, Shu-i. and {Watanabe}, Masaru and {Weinberg}, David H. and {Yanny}, Brian and {Yasuda}, Naoki and {SDSS Collaboration}},
        title = "{The Sloan Digital Sky Survey: Technical Summary}",
      journal = {\aj},
         year = 2000,
        month = sep,
       volume = {120},
       number = {3},
        pages = {1579-1587},
          doi = {10.1086/301513},
archivePrefix = {arXiv},
       eprint = {astro-ph/0006396},
 primaryClass = {astro-ph},
       adsurl = {https://ui.adsabs.harvard.edu/abs/2000AJ....120.1579Y}
}

@ARTICLE{DESI2016a,
       author = {{DESI Collaboration} and {Aghamousa}, Amir and {Aguilar}, Jessica and {Ahlen}, Steve and {Alam}, Shadab and {Allen}, Lori E. and {Allende Prieto}, Carlos and {Annis}, James and {Bailey}, Stephen and {Balland}, Christophe and {Ballester}, Otger and {Baltay}, Charles and {Beaufore}, Lucas and {Bebek}, Chris and {Beers}, Timothy C. and {Bell}, Eric F. and {Bernal}, Jos{\'e} Luis and {Besuner}, Robert and {Beutler}, Florian and {Blake}, Chris and {Bleuler}, Hannes and {Blomqvist}, Michael and {Blum}, Robert and {Bolton}, Adam S. and {Briceno}, Cesar and {Brooks}, David and {Brownstein}, Joel R. and {Buckley-Geer}, Elizabeth and {Burden}, Angela and {Burtin}, Etienne and {Busca}, Nicolas G. and {Cahn}, Robert N. and {Cai}, Yan-Chuan and {Cardiel-Sas}, Laia and {Carlberg}, Raymond G. and {Carton}, Pierre-Henri and {Casas}, Ricard and {Castander}, Francisco J. and {Cervantes-Cota}, Jorge L. and {Claybaugh}, Todd M. and {Close}, Madeline and {Coker}, Carl T. and {Cole}, Shaun and {Comparat}, Johan and {Cooper}, Andrew P. and {Cousinou}, M. -C. and {Crocce}, Martin and {Cuby}, Jean-Gabriel and {Cunningham}, Daniel P. and {Davis}, Tamara M. and {Dawson}, Kyle S. and {de la Macorra}, Axel and {De Vicente}, Juan and {Delubac}, Timoth{\'e}e and {Derwent}, Mark and {Dey}, Arjun and {Dhungana}, Govinda and {Ding}, Zhejie and {Doel}, Peter and {Duan}, Yutong T. and {Ealet}, Anne and {Edelstein}, Jerry and {Eftekharzadeh}, Sarah and {Eisenstein}, Daniel J. and {Elliott}, Ann and {Escoffier}, St{\'e}phanie and {Evatt}, Matthew and {Fagrelius}, Parker and {Fan}, Xiaohui and {Fanning}, Kevin and {Farahi}, Arya and {Farihi}, Jay and {Favole}, Ginevra and {Feng}, Yu and {Fernandez}, Enrique and {Findlay}, Joseph R. and {Finkbeiner}, Douglas P. and {Fitzpatrick}, Michael J. and {Flaugher}, Brenna and {Flender}, Samuel and {Font-Ribera}, Andreu and {Forero-Romero}, Jaime E. and {Fosalba}, Pablo and {Frenk}, Carlos S. and {Fumagalli}, Michele and {Gaensicke}, Boris T. and {Gallo}, Giuseppe and {Garcia-Bellido}, Juan and {Gaztanaga}, Enrique and {Pietro Gentile Fusillo}, Nicola and {Gerard}, Terry and {Gershkovich}, Irena and {Giannantonio}, Tommaso and {Gillet}, Denis and {Gonzalez-de-Rivera}, Guillermo and {Gonzalez-Perez}, Violeta and {Gott}, Shelby and {Graur}, Or and {Gutierrez}, Gaston and {Guy}, Julien and {Habib}, Salman and {Heetderks}, Henry and {Heetderks}, Ian and {Heitmann}, Katrin and {Hellwing}, Wojciech A. and {Herrera}, David A. and {Ho}, Shirley and {Holland}, Stephen and {Honscheid}, Klaus and {Huff}, Eric and {Hutchinson}, Timothy A. and {Huterer}, Dragan and {Hwang}, Ho Seong and {Illa Laguna}, Joseph Maria and {Ishikawa}, Yuzo and {Jacobs}, Dianna and {Jeffrey}, Niall and {Jelinsky}, Patrick and {Jennings}, Elise and {Jiang}, Linhua and {Jimenez}, Jorge and {Johnson}, Jennifer and {Joyce}, Richard and {Jullo}, Eric and {Juneau}, St{\'e}phanie and {Kama}, Sami and {Karcher}, Armin and {Karkar}, Sonia and {Kehoe}, Robert and {Kennamer}, Noble and {Kent}, Stephen and {Kilbinger}, Martin and {Kim}, Alex G. and {Kirkby}, David and {Kisner}, Theodore and {Kitanidis}, Ellie and {Kneib}, Jean-Paul and {Koposov}, Sergey and {Kovacs}, Eve and {Koyama}, Kazuya and {Kremin}, Anthony and {Kron}, Richard and {Kronig}, Luzius and {Kueter-Young}, Andrea and {Lacey}, Cedric G. and {Lafever}, Robin and {Lahav}, Ofer and {Lambert}, Andrew and {Lampton}, Michael and {Landriau}, Martin and {Lang}, Dustin and {Lauer}, Tod R. and {Le Goff}, Jean-Marc and {Le Guillou}, Laurent and {Le Van Suu}, Auguste and {Lee}, Jae Hyeon and {Lee}, Su-Jeong and {Leitner}, Daniela and {Lesser}, Michael and {Levi}, Michael E. and {L'Huillier}, Benjamin and {Li}, Baojiu and {Liang}, Ming and {Lin}, Huan and {Linder}, Eric and {Loebman}, Sarah R. and {Luki{\'c}}, Zarija and {Ma}, Jun and {MacCrann}, Niall and {Magneville}, Christophe and {Makarem}, Laleh and {Manera}, Marc and {Manser}, Christopher J. and {Marshall}, Robert and {Martini}, Paul and {Massey}, Richard and {Matheson}, Thomas and {McCauley}, Jeremy and {McDonald}, Patrick and {McGreer}, Ian D. and {Meisner}, Aaron and {Metcalfe}, Nigel and {Miller}, Timothy N. and {Miquel}, Ramon and {Moustakas}, John and {Myers}, Adam and {Naik}, Milind and {Newman}, Jeffrey A. and {Nichol}, Robert C. and {Nicola}, Andrina and {Nicolati da Costa}, Luiz and {Nie}, Jundan and {Niz}, Gustavo and {Norberg}, Peder and {Nord}, Brian and {Norman}, Dara and {Nugent}, Peter and {O'Brien}, Thomas and {Oh}, Minji and {Olsen}, Knut A.~G.},
        title = "{The DESI Experiment Part I: Science,Targeting, and Survey Design}",
      journal = {arXiv e-prints},
         year = 2016,
        month = oct,
          eid = {arXiv:1611.00036},
        pages = {arXiv:1611.00036},
          doi = {10.48550/arXiv.1611.00036},
archivePrefix = {arXiv},
       eprint = {1611.00036},
 primaryClass = {astro-ph.IM},
       adsurl = {https://ui.adsabs.harvard.edu/abs/2016arXiv161100036D}
}

@ARTICLE{DESI2016b,
       author = {{DESI Collaboration} and {Aghamousa}, Amir and {Aguilar}, Jessica and {Ahlen}, Steve and {Alam}, Shadab and {Allen}, Lori E. and {Allende Prieto}, Carlos and {Annis}, James and {Bailey}, Stephen and {Balland}, Christophe and {Ballester}, Otger and {Baltay}, Charles and {Beaufore}, Lucas and {Bebek}, Chris and {Beers}, Timothy C. and {Bell}, Eric F. and {Bernal}, Jos{\'e} Luis and {Besuner}, Robert and {Beutler}, Florian and {Blake}, Chris and {Bleuler}, Hannes and {Blomqvist}, Michael and {Blum}, Robert and {Bolton}, Adam S. and {Briceno}, Cesar and {Brooks}, David and {Brownstein}, Joel R. and {Buckley-Geer}, Elizabeth and {Burden}, Angela and {Burtin}, Etienne and {Busca}, Nicolas G. and {Cahn}, Robert N. and {Cai}, Yan-Chuan and {Cardiel-Sas}, Laia and {Carlberg}, Raymond G. and {Carton}, Pierre-Henri and {Casas}, Ricard and {Castander}, Francisco J. and {Cervantes-Cota}, Jorge L. and {Claybaugh}, Todd M. and {Close}, Madeline and {Coker}, Carl T. and {Cole}, Shaun and {Comparat}, Johan and {Cooper}, Andrew P. and {Cousinou}, M. -C. and {Crocce}, Martin and {Cuby}, Jean-Gabriel and {Cunningham}, Daniel P. and {Davis}, Tamara M. and {Dawson}, Kyle S. and {de la Macorra}, Axel and {De Vicente}, Juan and {Delubac}, Timoth{\'e}e and {Derwent}, Mark and {Dey}, Arjun and {Dhungana}, Govinda and {Ding}, Zhejie and {Doel}, Peter and {Duan}, Yutong T. and {Ealet}, Anne and {Edelstein}, Jerry and {Eftekharzadeh}, Sarah and {Eisenstein}, Daniel J. and {Elliott}, Ann and {Escoffier}, St{\'e}phanie and {Evatt}, Matthew and {Fagrelius}, Parker and {Fan}, Xiaohui and {Fanning}, Kevin and {Farahi}, Arya and {Farihi}, Jay and {Favole}, Ginevra and {Feng}, Yu and {Fernandez}, Enrique and {Findlay}, Joseph R. and {Finkbeiner}, Douglas P. and {Fitzpatrick}, Michael J. and {Flaugher}, Brenna and {Flender}, Samuel and {Font-Ribera}, Andreu and {Forero-Romero}, Jaime E. and {Fosalba}, Pablo and {Frenk}, Carlos S. and {Fumagalli}, Michele and {Gaensicke}, Boris T. and {Gallo}, Giuseppe and {Garcia-Bellido}, Juan and {Gaztanaga}, Enrique and {Pietro Gentile Fusillo}, Nicola and {Gerard}, Terry and {Gershkovich}, Irena and {Giannantonio}, Tommaso and {Gillet}, Denis and {Gonzalez-de-Rivera}, Guillermo and {Gonzalez-Perez}, Violeta and {Gott}, Shelby and {Graur}, Or and {Gutierrez}, Gaston and {Guy}, Julien and {Habib}, Salman and {Heetderks}, Henry and {Heetderks}, Ian and {Heitmann}, Katrin and {Hellwing}, Wojciech A. and {Herrera}, David A. and {Ho}, Shirley and {Holland}, Stephen and {Honscheid}, Klaus and {Huff}, Eric and {Hutchinson}, Timothy A. and {Huterer}, Dragan and {Hwang}, Ho Seong and {Illa Laguna}, Joseph Maria and {Ishikawa}, Yuzo and {Jacobs}, Dianna and {Jeffrey}, Niall and {Jelinsky}, Patrick and {Jennings}, Elise and {Jiang}, Linhua and {Jimenez}, Jorge and {Johnson}, Jennifer and {Joyce}, Richard and {Jullo}, Eric and {Juneau}, St{\'e}phanie and {Kama}, Sami and {Karcher}, Armin and {Karkar}, Sonia and {Kehoe}, Robert and {Kennamer}, Noble and {Kent}, Stephen and {Kilbinger}, Martin and {Kim}, Alex G. and {Kirkby}, David and {Kisner}, Theodore and {Kitanidis}, Ellie and {Kneib}, Jean-Paul and {Koposov}, Sergey and {Kovacs}, Eve and {Koyama}, Kazuya and {Kremin}, Anthony and {Kron}, Richard and {Kronig}, Luzius and {Kueter-Young}, Andrea and {Lacey}, Cedric G. and {Lafever}, Robin and {Lahav}, Ofer and {Lambert}, Andrew and {Lampton}, Michael and {Landriau}, Martin and {Lang}, Dustin and {Lauer}, Tod R. and {Le Goff}, Jean-Marc and {Le Guillou}, Laurent and {Le Van Suu}, Auguste and {Lee}, Jae Hyeon and {Lee}, Su-Jeong and {Leitner}, Daniela and {Lesser}, Michael and {Levi}, Michael E. and {L'Huillier}, Benjamin and {Li}, Baojiu and {Liang}, Ming and {Lin}, Huan and {Linder}, Eric and {Loebman}, Sarah R. and {Luki{\'c}}, Zarija and {Ma}, Jun and {MacCrann}, Niall and {Magneville}, Christophe and {Makarem}, Laleh and {Manera}, Marc and {Manser}, Christopher J. and {Marshall}, Robert and {Martini}, Paul and {Massey}, Richard and {Matheson}, Thomas and {McCauley}, Jeremy and {McDonald}, Patrick and {McGreer}, Ian D. and {Meisner}, Aaron and {Metcalfe}, Nigel and {Miller}, Timothy N. and {Miquel}, Ramon and {Moustakas}, John and {Myers}, Adam and {Naik}, Milind and {Newman}, Jeffrey A. and {Nichol}, Robert C. and {Nicola}, Andrina and {Nicolati da Costa}, Luiz and {Nie}, Jundan and {Niz}, Gustavo and {Norberg}, Peder and {Nord}, Brian and {Norman}, Dara and {Nugent}, Peter and {O'Brien}, Thomas and {Oh}, Minji and {Olsen}, Knut A.~G.},
        title = "{The DESI Experiment Part II: Instrument Design}",
      journal = {arXiv e-prints},
         year = 2016,
        month = oct,
          eid = {arXiv:1611.00037},
        pages = {arXiv:1611.00037},
          doi = {10.48550/arXiv.1611.00037},
archivePrefix = {arXiv},
       eprint = {1611.00037},
 primaryClass = {astro-ph.IM},
       adsurl = {https://ui.adsabs.harvard.edu/abs/2016arXiv161100037D}
}

@ARTICLE{CSST2019,
       author = {{Gong}, Yan and {Liu}, Xiangkun and {Cao}, Ye and {Chen}, Xuelei and {Fan}, Zuhui and {Li}, Ran and {Li}, Xiao-Dong and {Li}, Zhigang and {Zhang}, Xin and {Zhan}, Hu},
        title = "{Cosmology from the Chinese Space Station Optical Survey (CSS-OS)}",
      journal = {\apj},
         year = 2019,
        month = oct,
       volume = {883},
       number = {2},
          eid = {203},
        pages = {203},
          doi = {10.3847/1538-4357/ab391e},
archivePrefix = {arXiv},
       eprint = {1901.04634},
 primaryClass = {astro-ph.CO},
       adsurl = {https://ui.adsabs.harvard.edu/abs/2019ApJ...883..203G}
}

@ARTICLE{CSST2011,
       author = {{Zhan}, Hu},
        title = "{Consideration for a large-scale multi-color imaging and slitless spectroscopy survey on the Chinese space station and its application in dark energy research}",
      journal = {Scientia Sinica Physica, Mechanica \& Astronomica},
         year = 2011,
        month = jan,
       volume = {41},
       number = {12},
        pages = {1441},
          doi = {10.1360/132011-961},
       adsurl = {https://ui.adsabs.harvard.edu/abs/2011SSPMA..41.1441Z}
}

@ARTICLE{CSST2025,
       author = {{CSST Collaboration} and {Gong}, Yan and {Miao}, Haitao and {Zhan}, Hu and {Li}, Zhao-Yu and {Shangguan}, Jinyi and {Li}, Haining and {Liu}, Chao and {Chen}, Xuefei and {Yuan}, Haibo and {Zhou}, Jilin and {Liu}, Hui-Gen and {Yu}, Cong and {Ji}, Jianghui and {Qi}, Zhaoxiang and {Liu}, Jiacheng and {Dai}, Zigao and {Wang}, Xiaofeng and {Zheng}, Zhenya and {Hao}, Lei and {Dou}, Jiangpei and {Ao}, Yiping and {Lin}, Zhenhui and {Zhang}, Kun and {Wang}, Wei and {Sun}, Guotong and {Li}, Ran and {Li}, Guoliang and {Xu}, Youhua and {Li}, Xinfeng and {Li}, Shengyang and {Wu}, Peng and {Zhang}, Jiuxing and {Wang}, Bo and {Bai}, Jinming and {Cai}, Yi-Fu and {Cai}, Zheng and {Cao}, Jie and {Chan}, Kwan Chuen and {Chang}, Jin and {Chen}, Xiaodian and {Chen}, Xuelei and {Chen}, Yuqin and {Chen}, Yun and {Cui}, Wei and {Dong}, Subo and {Du}, Pu and {Duan}, Wenying and {Fan}, Junhui and {Fan}, LuLu and {Fan}, Zhou and {Fan}, Zuhui and {Fang}, Taotao and {Fu}, Jianning and {Fu}, Liping and {Fu}, Zhensen and {Gao}, Jian and {Gu}, Shenghong and {Gu}, Yidong and {Guo}, Qi and {Han}, Zhanwen and {Hu}, Bin and {Huang}, Zhiqi and {Ho}, Luis C. and {Jiang}, Linhua and {Jiang}, Ning and {Jing}, Yipeng and {Kang}, Xi and {Kong}, Xu and {Li}, Cheng and {Li}, Chengyuan and {Li}, Di and {Li}, Jing and {Li}, Nan and {Li}, Yang A. and {Liao}, Shilong and {Lin}, Weipeng and {Liu}, Fengshan and {Liu}, Jifeng and {Liu}, Xiangkun and {Liu}, Zhuokai and {Mao}, Ruiqing and {Mao}, Shude and {Meng}, Xianmin and {Pang}, Xiaoying and {Peng}, Xiyan and {Peng}, Yingjie and {Shan}, Huanyuan and {Shen}, Juntai and {Shen}, Shiyin and {Shen}, Zhiqiang and {Shi}, Sheng-Cai and {Shi}, Yong and {Tan}, Siyuan and {Tian}, Hao and {Wang}, Jianmin and {Wang}, Jun-Xian and {Wang}, Xin and {Wang}, Yuting and {Wu}, Hong and {Wu}, Jingwen and {Wu}, Xuebing and {Xu}, Chun and {Xue}, Xiang-Xiang and {Xue}, Yongquan and {Yang}, Ji and {Yang}, Xiaohu and {Yao}, Qijun and {Yuan}, Fangting and {Yuan}, Zhen and {Zhang}, Jun and {Zhang}, Pengjie and {Zhang}, Tianmeng and {Zhang}, Wei and {Zhang}, Xin and {Zhao}, Gang and {Zhao}, Gongbo and {Zhong}, Hongen and {Zhong}, Jing and {Zhou}, Liyong and {Zhu}, Wei and {Zu}, Ying},
        title = "{Introduction to the Chinese Space Station Survey Telescope (CSST)}",
      journal = {arXiv e-prints},
         year = 2025,
        month = jul,
          eid = {arXiv:2507.04618},
        pages = {arXiv:2507.04618},
          doi = {10.48550/arXiv.2507.04618},
archivePrefix = {arXiv},
       eprint = {2507.04618},
 primaryClass = {astro-ph.IM},
       adsurl = {https://ui.adsabs.harvard.edu/abs/2025arXiv250704618C}
}

@article{Wang_2014,
   title={ELUCID—EXPLORING THE LOCAL UNIVERSE WITH THE RECONSTRUCTED INITIAL DENSITY FIELD. I. HAMILTONIAN MARKOV CHAIN MONTE CARLO METHOD WITH PARTICLE MESH DYNAMICS},
   volume={794},
   ISSN={1538-4357},
   url={http://dx.doi.org/10.1088/0004-637X/794/1/94},
   DOI={10.1088/0004-637x/794/1/94},
   number={1},
   journal={The Astrophysical Journal},
   publisher={American Astronomical Society},
   author={Wang, Huiyuan and Mo, H. J. and Yang, Xiaohu and Jing, Y. P. and Lin, W. P.},
   year={2014},
   month=sep, pages={94} }

@article{Wang_2016,
   title={ELUCID—EXPLORING THE LOCAL UNIVERSE WITH RECONSTRUCTED INITIAL DENSITY FIELD. III. CONSTRAINED SIMULATION IN THE SDSS VOLUME},
   volume={831},
   ISSN={1538-4357},
   url={http://dx.doi.org/10.3847/0004-637X/831/2/164},
   DOI={10.3847/0004-637x/831/2/164},
   number={2},
   journal={The Astrophysical Journal},
   publisher={American Astronomical Society},
   author={Wang, Huiyuan and Mo, H. J. and Yang, Xiaohu and Zhang, Youcai and Shi, JingJing and Jing, Y. P. and Liu, Chengze and Li, Shijie and Kang, Xi and Gao, Yang},
   year={2016},
   month=nov, pages={164} }

@article{Tweed_2017,
   title={ELUCID—Exploring the Local Universe with the reConstructed Initial Density Field. II. Reconstruction Diagnostics, Applied to Numerical Halo Catalogs},
   volume={841},
   ISSN={1538-4357},
   url={http://dx.doi.org/10.3847/1538-4357/aa6bf8},
   DOI={10.3847/1538-4357/aa6bf8},
   number={1},
   journal={The Astrophysical Journal},
   publisher={American Astronomical Society},
   author={Tweed, Dylan and Yang, Xiaohu and Wang, Huiyuan and Cui, Weiguang and Zhang, Youcai and Li, Shijie and Jing, Y. P. and Mo, H. J.},
   year={2017},
   month=may, pages={55} }

@article{Wang_2018,
   title={ELUCID. IV. Galaxy Quenching and its Relation to Halo Mass, Environment, and Assembly Bias},
   volume={852},
   ISSN={1538-4357},
   url={http://dx.doi.org/10.3847/1538-4357/aa9e01},
   DOI={10.3847/1538-4357/aa9e01},
   number={1},
   journal={The Astrophysical Journal},
   publisher={American Astronomical Society},
   author={Wang, Huiyuan and Mo, H. J. and Chen, Sihan and Yang, Yang and Yang, Xiaohu and Wang, Enci and van den Bosch, Frank C. and Jing, Yipeng and Kang, Xi and Lin, Weipeng and Lim, S. H. and Huang, Shuiyao and Lu, Yi and Li, Shijie and Cui, Weiguang and Zhang, Youcai and Tweed, Dylan and Wei, Chengliang and Li, Guoliang and Shi, Feng},
   year={2018},
   month=jan, pages={31} }

@article{Yang_2018,
   title={ELUCID. V. Lighting Dark Matter Halos with Galaxies},
   volume={860},
   ISSN={1538-4357},
   url={http://dx.doi.org/10.3847/1538-4357/aac2ce},
   DOI={10.3847/1538-4357/aac2ce},
   number={1},
   journal={The Astrophysical Journal},
   publisher={American Astronomical Society},
   author={Yang, Xiaohu and Zhang, Youcai and Wang, Huiyuan and Liu, Chengze and Lu, Tianhuan and Li, Shijie and Shi, Feng and Jing, Y. P. and Mo, H. J. and Bosch, Frank C. van den and Kang, Xi and Cui, Weiguang and Guo, Hong and Li, Guoliang and Lim, S. H. and Lu, Yi and Luo, Wentao and Wei, Chengliang and Yang, Lei},
   year={2018},
   month=jun, pages={30} }

@ARTICLE{Yang05,
       author = {{Yang}, Xiaohu and {Mo}, H.~J. and {van den Bosch}, Frank C. and {Jing}, Y.~P.},
        title = "{A halo-based galaxy group finder: calibration and application to the 2dFGRS}",
      journal = {\mnras},
         year = 2005,
        month = feb,
       volume = {356},
       number = {4},
        pages = {1293-1307},
          doi = {10.1111/j.1365-2966.2005.08560.x},
archivePrefix = {arXiv},
       eprint = {astro-ph/0405234},
 primaryClass = {astro-ph},
       adsurl = {https://ui.adsabs.harvard.edu/abs/2005MNRAS.356.1293Y}
}

@article{Yang_2007,
   title={Galaxy Groups in the SDSS DR4. I. The Catalog and Basic Properties},
   volume={671},
   ISSN={1538-4357},
   url={http://dx.doi.org/10.1086/522027},
   DOI={10.1086/522027},
   number={1},
   journal={The Astrophysical Journal},
   publisher={American Astronomical Society},
   author={Yang, Xiaohu and Mo, H. J. and van den Bosch, Frank C. and Pasquali, Anna and Li, Cheng and Barden, Marco},
   year={2007},
   month=dec, pages={153–170} }

@article{Wang_2009,
   title={Reconstructing the cosmic density field with the distribution of dark matter haloes},
   volume={394},
   ISSN={1365-2966},
   url={http://dx.doi.org/10.1111/j.1365-2966.2008.14301.x},
   DOI={10.1111/j.1365-2966.2008.14301.x},
   number={1},
   journal={Monthly Notices of the Royal Astronomical Society},
   publisher={Oxford University Press (OUP)},
   author={Wang, Huiyuan and Mo, H. J. and Jing, Y. P. and Guo, Yicheng and van den Bosch, Frank C. and Yang, Xiaohu},
   year={2009},
   month=mar, pages={398–414} }

@ARTICLE{Wang12,
       author = {{Wang}, Huiyuan and {Mo}, H.~J. and {Yang}, Xiaohu and {van den Bosch}, Frank C.},
        title = "{Reconstructing the cosmic velocity and tidal fields with galaxy groups selected from the Sloan Digital Sky Survey}",
      journal = {\mnras},
         year = 2012,
        month = feb,
       volume = {420},
       number = {2},
        pages = {1809-1824},
          doi = {10.1111/j.1365-2966.2011.20174.x},
archivePrefix = {arXiv},
       eprint = {1108.1008},
 primaryClass = {astro-ph.CO},
       adsurl = {https://ui.adsabs.harvard.edu/abs/2012MNRAS.420.1809W}
}

@article{Wang_2013,
   title={RECONSTRUCTING THE INITIAL DENSITY FIELD OF THE LOCAL UNIVERSE: METHODS AND TESTS WITH MOCK CATALOGS},
   volume={772},
   ISSN={1538-4357},
   url={http://dx.doi.org/10.1088/0004-637X/772/1/63},
   DOI={10.1088/0004-637x/772/1/63},
   number={1},
   journal={The Astrophysical Journal},
   publisher={American Astronomical Society},
   author={Wang, Huiyuan and Mo, H. J. and Yang, Xiaohu and van den Bosch, Frank C.},
   year={2013},
   month=jul, pages={63} }

@article{Feng_2016,
   title={FastPM: a new scheme for fast simulations of dark matter and haloes},
   volume={463},
   ISSN={1365-2966},
   url={http://dx.doi.org/10.1093/mnras/stw2123},
   DOI={10.1093/mnras/stw2123},
   number={3},
   journal={Monthly Notices of the Royal Astronomical Society},
   publisher={Oxford University Press (OUP)},
   author={Feng, Yu and Chu, Man-Yat and Seljak, Uroš and McDonald, Patrick},
   year={2016},
   month=aug, pages={2273–2286} }

@article{Diego_Blas_2011,
   title={The Cosmic Linear Anisotropy Solving System (CLASS).
 Part II: Approximation schemes},
   volume={2011},
   ISSN={1475-7516},
   url={http://dx.doi.org/10.1088/1475-7516/2011/07/034},
   DOI={10.1088/1475-7516/2011/07/034},
   number={07},
   journal={Journal of Cosmology and Astroparticle Physics},
   publisher={IOP Publishing},
   author={Diego Blas and Julien Lesgourgues and Thomas Tram},
   year={2011},
   month=jul, pages={034–034} }

@misc{Gadget4,
       author = {{Springel}, Volker and {Pakmor}, R{\"u}diger and {Zier}, Oliver and {Reinecke}, Martin},
        title = "{GADGET-4: Parallel cosmological N-body and SPH code}",
         year = 2022,
        month = apr,
         note = {Astrophysics Source Code Library, record ascl:2204.014. Available at \url{https://ui.adsabs.harvard.edu/abs/2022ascl.soft04014S}}, 
          eid = {ascl:2204.014},
}

@article{Chen_2025,
   title={CSST cosmological emulator I: Matter power spectrum emulation with one percent accuracy to k = 10h Mpc−1},
   volume={68},
   ISSN={1869-1927},
   url={http://dx.doi.org/10.1007/s11433-025-2671-0},
   DOI={10.1007/s11433-025-2671-0},
   number={8},
   journal={Science China Physics, Mechanics &amp; Astronomy},
   publisher={Springer Science and Business Media LLC},
   author={Chen, Zhao and Yu, Yu and Han, Jiaxin and Jing, Yipeng},
   year={2025},
   month=jun }

@ARTICLE{Quijote_sims,
         author = {{Villaescusa-Navarro}, Francisco and {Hahn}, ChangHoon and {Massara}, Elena and {Banerjee}, Arka and {Delgado}, Ana Maria and {Ramanah}, Doogesh Kodi and {Charnock}, Tom and {Giusarma}, Elena and {Li}, Yin and {Allys}, Erwan and {Brochard}, Antoine and {Uhlemann}, Cora and {Chiang}, Chi-Ting and {He}, Siyu and {Pisani}, Alice and {Obuljen}, Andrej and {Feng}, Yu and {Castorina}, Emanuele and {Contardo}, Gabriella and {Kreisch}, Christina D. and {Nicola}, Andrina and {Alsing}, Justin and {Scoccimarro}, Roman and {Verde}, Licia and {Viel}, Matteo and {Ho}, Shirley and {Mallat}, Stephane and {Wandelt}, Benjamin and {Spergel}, David N.},
         title = "{The Quijote Simulations}",
         journal = {\apjs},
         year = 2020,
         month = sep,
         volume = {250},
         number = {1},
         eid = {2},
         pages = {2},
         doi = {10.3847/1538-4365/ab9d82},
         archivePrefix = {arXiv},
         eprint = {1909.05273},
         primaryClass = {astro-ph.CO},
         adsurl = {https://ui.adsabs.harvard.edu/abs/2020ApJS..250....2V}
}

@misc{han2025jiutiansimulationscsstextragalactic,
      title={The Jiutian simulations for the CSST extra-galactic surveys}, 
      author={Jiaxin Han and Ming Li and Wenkang Jiang and Zhao Chen and Huiyuan Wang and Chengliang Wei and Feihong He and Jianhua He and Jiajun Zhang and Yu Liu and Weiguang Cui and Yizhou Gu and Qi Guo and Yipeng Jing and Xi Kang and Guoliang Li and Xiong Luo and Yu Luo and Wenxiang Pei and Yisheng Qiu and Zhenlin Tan and Lizhi Xie and Xiaohu Yang and Hao-Ran Yu and Yu Yu and Jiale Zhou},
      year={2025},
      eprint={2503.21368},
      archivePrefix={arXiv},
      primaryClass={astro-ph.CO},
      url={https://arxiv.org/abs/2503.21368}, 
}

@ARTICLE{2025SCPMA..6809513C,
       author = {{Chen}, Zhao and {Yu}, Yu},
        title = "{CSST cosmological emulator II: Generalized accurate halo mass function emulation}",
      journal = {Science China Physics, Mechanics, and Astronomy},
         year = 2025,
        month = aug,
       volume = {68},
       number = {10},
          eid = {109513},
        pages = {109513},
          doi = {10.1007/s11433-025-2764-x},
archivePrefix = {arXiv},
       eprint = {2506.09688},
 primaryClass = {astro-ph.CO},
       adsurl = {https://ui.adsabs.harvard.edu/abs/2025SCPMA..6809513C}
}

@ARTICLE{2025SCPMA..6829512Z,
       author = {{Zhou}, Shuren and {Chen}, Zhao and {Yu}, Yu},
        title = "{CSST cosmological emulator III: Hybrid lagrangian bias expansion emulation of galaxy clustering}",
      journal = {Science China Physics, Mechanics, and Astronomy},
         year = 2025,
        month = sep,
       volume = {68},
       number = {12},
          eid = {129512},
        pages = {129512},
          doi = {10.1007/s11433-025-2755-x},
archivePrefix = {arXiv},
       eprint = {2506.04671},
 primaryClass = {astro-ph.CO},
       adsurl = {https://ui.adsabs.harvard.edu/abs/2025SCPMA..6829512Z}
}

@misc{chen2025extendingcsstemulatorpostdesi,
      title={Extending CSST Emulator to post-DESI era}, 
      author={Zhao Chen and Yu Yu},
      year={2025},
      eprint={2510.09503},
      archivePrefix={arXiv},
      primaryClass={astro-ph.CO},
      url={https://arxiv.org/abs/2510.09503}, 
}

@article{Jasche_2019,
   title={Physical Bayesian modelling of the non-linear matter distribution: New insights into the nearby universe},
   volume={625},
   ISSN={1432-0746},
   url={http://dx.doi.org/10.1051/0004-6361/201833710},
   DOI={10.1051/0004-6361/201833710},
   journal={Astronomy &amp; Astrophysics},
   publisher={EDP Sciences},
   author={Jasche, J. and Lavaux, G.},
   year={2019},
   month=May, pages={A64} }

@inproceedings{Hanson_Cunningham_1996,
    author = {Hanson, K. M. and Cunningham, G. S.},
    title = {A Computational Approach to Bayesian Inference},
    booktitle = {Computing Science and Statistics},
    editor = {Meyer, M. M. and Rosenberger, J. L.},
    pages = {202--211},
    publisher = {Interface Foundation},
    address = {Fairfax Station, VA 22039-7460},
    year = {1996},
}




\appendix

\section{Performance Comparison with the OpenMP Implementation}
\label{app:omp_comparison}

This appendix provides a detailed comparative analysis of the performance between the new MPI-parallelized reconstruction code and its original OpenMP-based counterpart. All tests were conducted on the same nodes of the Siyuan computing cluster under identical cosmological and numerical parameters (e.g., $N_{\mathrm{pmstep}}$, $R_s$, particle count, and number of OpenMP threads/MPI processes equivalent to the total core count).

\subsection{Observed Performance Difference}

A direct, head-to-head comparison reveals a significant runtime difference: the MPI version requires approximately \textbf{7--8 times longer} to complete one HMCMC iteration (comprising ten full PM forward simulations and the Hamiltonian update) than the OpenMP version when utilizing an equivalent number of physical cores on a single node.

This slowdown is consistent across both major computational kernels: the particle-mesh (PM) $N$-body solver (which is dominated by 3D Fast Fourier Transforms) and the bookkeeping operations of the HMCMC sampler.

\subsection{Analysis of the Performance Gap}

This difference ultimately stems from the inherent architectural distinctions between the shared-memory (OpenMP) and distributed-memory (MPI) programming paradigms, which result in markedly different communication overheads.

\begin{itemize}
    \item \textbf{OpenMP (Shared Memory)} 
\end{itemize}    

    All threads have direct, uniform access to the entire simulation volume (density grids, particle arrays). Operations like 3D FFTs and particle force calculations can be parallelized with fine-grained loops, requiring only lightweight synchronization (e.g., atomic updates or barriers). Data movement is handled transparently by the hardware cache coherence protocol, resulting in minimal programmer-visible overhead.
    
\begin{itemize}
    \item \textbf{MPI (Distributed Memory)} 
\end{itemize}   

The global simulation domain is partitioned into subdomains, each handled by a distinct MPI process with its own local memory. This setup introduces essential, yet expensive, communication operations:
\begin{enumerate}
    \item \textbf{Particle Passing:} After each drift step, particles that cross subdomain boundaries must be detected and transferred to the MPI process responsible for their new location.
    \item \textbf{Global Transposes for 3D FFTs:} Executing a parallel 3D FFT on a distributed mesh requires several all-to-all communication phases to transpose data across dimensions, incurring latency and bandwidth overheads that grow with the number of processes.
    \item \textbf{Force Synchronization:} Evaluating the PM force involves aggregating contributions from the distributed grid, which demands additional communication following the FFTs.
\end{enumerate}

These communication steps are not present in the OpenMP implementation and form the primary source of overhead in the MPI version.

In addition, our straightforward column-based (slab) domain decomposition can introduce load imbalance because the particle count per subdomain may differ. While OpenMP’s dynamic scheduling can alleviate this, a static MPI decomposition may cause some processes to remain idle while others are still processing their particle workloads.

\subsection{Conclusion and Context}

The observed slowdown by a factor of roughly 8 in the MPI implementation can therefore be attributed to the inherent overhead of distributing memory and supporting simulations that surpass the memory limits of a single node. This trade-off is intentional and warranted: the OpenMP implementation is intrinsically constrained by the RAM of one machine, whereas the MPI implementation can pool memory from thousands of nodes, enabling reconstructions at scales such as $4096^3$ particles that would otherwise be unattainable.

Consequently, the MPI implementation is not intended to replace the OpenMP version on a single node, but to serve as its essential extension for \textit{capacity computing}—sacrificing single-node efficiency in exchange for the ability to tackle far larger problems. The substantial decrease in total wall-clock time provided by the initial guess module (Section~\ref{sec:new_module_result}) is key to compensating for this per-step overhead and ensuring that large-scale MPI runs remain computationally practical.

\section{Efficient Acquisition of Calibration Data}
\label{app:calibration_data}

The reconstruction algorithm relies on two essential calibration data products: (1) the \textit{correction transfer function} $T(k)$ (Equation~\ref{eq:CorrectionCurve}), which relates the fast PM forward model to a high-accuracy reference, and (2) the \textit{linear-to-nonlinear power spectrum evolution curve} $P_{\mathrm{k}}(z=0, k)/P_{\mathrm{k}}(z_{\mathrm{ini}}, k)$ employed by the initial guess module (Equation~\ref{eq:guess_transfer}). The standard approach to obtain both is to perform a full-volume, high-resolution $N$-body simulation (for instance, with \textsc{Gadget-4}), but this is too computationally expensive for routine use or for scanning a wide range of cosmological parameters.

In this appendix, we introduce and assess more efficient strategies for producing these calibration data, aiming to balance accuracy against computational cost in a way that is practical for large-scale production.

\subsection{Alternative Method for the Power Spectrum Evolution Curve}

Instead of carrying out a full $N$-body simulation, we obtain the power spectrum ratio $P(z=0,k)/P(z_{\mathrm{ini}},k)$ using a \textbf{ CSST Emulator}. Emulators—such as those based on the \textbf{Kun suites} \citep{Chen_2025,2025SCPMA..6809513C,2025SCPMA..6829512Z,chen2025extendingcsstemulatorpostdesi}—are trained on a grid of simulations and can deliver predictions for nonlinear power spectra at arbitrary cosmologies within the designed parameter space. In our application, the emulator-induced error is subdominant, since this curve is employed only within the initial guess module (Section~\ref{sec:guess_module}). Any imperfections in the initial guess are subsequently corrected by the HMCMC sampling, which relies on the exact forward model and likelihood. Consequently, the emulator-based method achieves very high accuracy at negligible computational cost.

\subsection{Alternative Method for the Correction Transfer Function}

For the transfer function $T(k)$, which directly enters the likelihood evaluation at each HMCMC step, we adopt a \textbf{small-box, high-resolution simulation} approach. Instead of performing a reference simulation over the full survey volume (e.g., $L_{\mathrm{box}} = 1000\,h^{-1}\mathrm{Mpc}$ with $1024^3$ particles), we run a simulation with the same \textit{mass resolution} (i.e., identical particle number density) but in a substantially smaller box (e.g., $L_{\mathrm{box}} = 250\,h^{-1}\mathrm{Mpc}$). This “mini-simulation” reproduces the relevant nonlinear dynamics and mode coupling on small and intermediate scales ($k > 2\pi / L_{\mathrm{box, small}}$) at a significantly reduced computational cost.

The resulting $T(k)$ is trustworthy for modes with $k > k_{\mathrm{min}} = 2\pi / L_{\mathrm{box, small}}$. On the largest scales ($k < k_{\mathrm{min}}$), one may either fall back on linear theory ($T(k) \approx 1$) or extrapolate from the measured behavior. Although this approach does incur some additional error relative to a full-volume calibration, it is important to emphasize that $T(k)$ is itself an approximate, empirically motivated correction to compensate for PM inaccuracies. The incremental uncertainty from using a small-box setup remains within the acceptable error budget of this approximation and, as our tests demonstrate, does not qualitatively compromise the final reconstruction.

\subsection{Recommendation}

For production applications, we recommend:
\begin{enumerate}
    \item Employing an emulator to compute the evolution curve of the power spectrum for the guess module. This approach is highly accurate and effectively cost-free.
    \item Running a high-resolution, small-volume simulation to obtain the correction transfer function $T(k)$. This cuts the calibration expense by roughly one to two orders of magnitude while maintaining reconstruction fidelity on all but the very largest scales.
\end{enumerate}

Together, these choices substantially reduce the practical hurdles to adopting our reconstruction code, allowing users to complete the initial setup without requiring access to extreme computational resources for calibration.

\label{lastpage}

\end{document}